\documentclass[journal]{IEEEtran}
\ifCLASSINFOpdf
\else
\fi
\usepackage{graphicx}
\graphicspath{{FiguresPDF/}}
\usepackage{multicol}
\usepackage{subcaption}
\usepackage{caption}
\usepackage{amsfonts}
\usepackage{comment}
\usepackage{xcolor}

\usepackage{soul}

\usepackage{algorithm,algpseudocode}

\usepackage{amsmath}
\usepackage{cleveref}

\algrenewcommand\textproc{}

\usepackage[most]{tcolorbox}
\usepackage{authblk}

\begin{document}
%
% paper title
% Titles are generally capitalized except for words such as a, an, and, as,
% at, but, by, for, in, nor, of, on, or, the, to and up, which are usually
% not capitalized unless they are the first or last word of the title.
% Linebreaks \\ can be used within to get better formatting as desired.
% Do not put math or special symbols in the title.
\title{Distributed CNN Inference on Resource-Constrained UAVs for Surveillance Systems: Design and Optimization}
%
%
% author names and IEEE memberships
% note positions of commas and nonbreaking spaces ( ~ ) LaTeX will not break
% a structure at a ~ so this keeps an author's name from being broken across
% two lines.
% use \thanks{} to gain access to the first footnote area
% a separate \thanks must be used for each paragraph as LaTeX2e's \thanks
% was not built to handle multiple paragraphs
%

\author{Mohammed Jouhari,
Abdulla Al-Ali,
Emna Baccour, 
Amr Mohamed,
Aiman Erbad,\\
\vspace{-3mm} 
Mohsen Guizani,
Mounir Hamdi
\thanks{Corresponding author: Mohammed Jouhari (email: m.jouhari@qu.edu.qa).}
\thanks{M. Jouhari, A. Al-Ali, A. Mohamed and M. Guizani are with CSE department, College of Engineering, Qatar University, Qatar (email: m.jouhari, abdulla.alali, amrm, mguizani@qu.edu.qa).}
\thanks{E. Baccour, A. Erbad and M. Hamdi are with College of Science and Engineering, Hamad Bin Khalifa University, Qatar (email: EBaccourEpBesaid, AErbad, mhamdi@hbku.edu.qa).}
\thanks{}
}

% note the % following the last \IEEEmembership and also \thanks - 
% these prevent an unwanted space from occurring between the last author name
% and the end of the author line. i.e., if you had this:
% 
% \author{....lastname \thanks{...} \thanks{...} }
%                     ^------------^------------^----Do not want these spaces!
%
% a space would be appended to the last name and could cause every name on that
% line to be shifted left slightly. This is one of those "LaTeX things". For
% instance, "\textbf{A} \textbf{B}" will typeset as "A B" not "AB". To get
% "AB" then you have to do: "\textbf{A}\textbf{B}"
% \thanks is no different in this regard, so shield the last } of each \thanks
% that ends a line with a % and do not let a space in before the next \thanks.
% Spaces after \IEEEmembership other than the last one are OK (and needed) as
% you are supposed to have spaces between the names. For what it is worth,
% this is a minor point as most people would not even notice if the said evil
% space somehow managed to creep in.

% The paper headers
\markboth{IEEE Internet of Things Journal}%
{Shell \MakeLowercase{\textit{et al.}}: Bare Demo of IEEEtran.cls for IEEE Journals}
% The only time the second header will appear is for the odd numbered pages
% after the title page when using the twoside option.
% 
% *** Note that you probably will NOT want to include the author's ***
% *** name in the headers of peer review papers.                   ***
% You can use \ifCLASSOPTIONpeerreview for conditional compilation here if
% you desire.

% If you want to put a publisher's ID mark on the page you can do it like
% this:
%\IEEEpubid{0000--0000/00\$00.00~\copyright~2015 IEEE}
% Remember, if you use this you must call \IEEEpubidadjcol in the second
% column for its text to clear the IEEEpubid mark.

% use for special paper notices
%\IEEEspecialpapernotice{(Invited Paper)}

% make the title area
\maketitle

% As a general rule, do not put math, special symbols or citations
% in the abstract or keywords.
\begin{abstract}
Unmanned Aerial Vehicles (UAVs) have attracted great interest in the last few years owing to their ability to cover large areas and access difficult and hazardous target zones, which is not the case of traditional systems relying on direct observations obtained from fixed cameras and sensors. Furthermore, thanks to the advancements in computer vision and machine learning, UAVs are being adopted for a broad range of solutions and applications. However, Deep Neural Networks (DNNs) are progressing toward deeper and complex models that prevent them from being executed on-board. In this paper, we propose a DNN distribution methodology within UAVs to enable data classification in resource-constrained devices and avoid extra delays introduced by the server-based solutions due to data communication over air-to-ground links.The proposed method is formulated as an optimization problem that aims to minimize the latency between data collection and decision-making while considering the mobility model and the resource constraints of the UAVs as part of the air-to-air communication. We also introduce the mobility prediction to adapt our system to the dynamics of UAVs and the network variation. The simulation conducted to evaluate the performance and benchmark the proposed methods, namely Optimal UAV-based Layer Distribution (OULD) and OULD with Mobility Prediction (OULD-MP), were run in an HPC cluster. The obtained results show that our optimization solution outperforms the existing and heuristic-based approaches.
\end{abstract}

% Note that keywords are not normally used for peerreview papers.
\begin{IEEEkeywords}
Unmanned Aerial Vehicles, Distributed Machine Learning, Deep Neural Networks, Convolutional Neural Network.
\end{IEEEkeywords}

% For peer review papers, you can put extra information on the cover
% page as needed:
% \ifCLASSOPTIONpeerreview
% \begin{center} \bfseries EDICS Category: 3-BBND \end{center}
% \fi
%
% For peerreview papers, this IEEEtran command inserts a page break and
% creates the second title. It will be ignored for other modes.
\IEEEpeerreviewmaketitle

\section{Introduction}
% The very first letter is a 2 line initial drop letter followed
% by the rest of the first word in caps.
% 
% form to use if the first word consists of a single letter:
% \IEEEPARstart{A}{demo} file is ....
% 
% form to use if you need the single drop letter followed by
% normal text (unknown if ever used by the IEEE):
% \IEEEPARstart{A}{}demo file is ....
% 
% Some journals put the first two words in caps:
% \IEEEPARstart{T}{his demo} file is ....
% 
% Here we have the typical use of a "T" for an initial drop letter
% and "HIS" in caps to complete the first word.
\IEEEPARstart{I}{n} the last decade, UAVs have been proposed as an alternative to the traditional technologies involved in a plethora of applications ranging from large-scale agriculture in search of weeds and pests \cite{7989347}, goods delivery \cite{8482480}, and wild life recording \cite{doi:10.1890/120150} to smart city monitoring \cite{8675178} and rescue operations \cite{8924547}. Meanwhile, technologies that have been used to remotely monitor large areas such as Satellite Remote Sensing (SRS) and Sensor Networks (SNs) are losing interest against UAVs \cite{PADRO2019130}. This revolutionary advancement in the monitoring systems is related to the maturity of UAVs and the challenges that both aforementioned technologies encounter, including the implementation cost, the non-ability to perform close monitoring for the case of SRS systems, the permanent deployment of sensors, and the need for constant human intervention and maintenance for the case of SNs. Hence, driven by the need for better scene coverage and prompt interventions in case of incidents, UAVs are proposed as a new innovative and cost-effective solution for low altitude sensing with zero deployments. Additionally, UAVs present a rapid and flexible data acquisition system that can provide close monitoring of human activity and objects from different angles and altitudes, resulting in high-resolution data used to enhance complex events detection. % 8207649

Such distinctive performance of drones encouraged the emergence of more critical and sophisticated missions in uncertain and potentially harsh environments, many of them was not even envisaged a couple of decades ago, including military border surveillance and oil/gas offshore inspection \cite{8337903}. Some of these solutions require more than one UAV to grant higher system reliability and scalability, and ensure dynamic and flexible surveillance \cite{8337901}. More specifically, the swarm of drones equipped with different types of sensors are distributed in the mission zone to gather the real-time data and report the on-site information to the command center in order to take immediate measures. Another major advantage of a UAV swarm is the ability to handle more complex operations by distributing and performing parallel tasks to reduce the execution time and guarantee a better fault tolerance \cite{8651796}. 

%In this context, a prominent example is the use of  cooperative UAVs to maximize the wireless coverage by acting as wireless base stations or relays \cite{8660516} or by leveraging flying ad hoc networking (FANET) advantages to mitigate the communication range restriction \cite{9044378}. 
%, 8950047 \cite{7463007} \cite{8939564}

%While UAV swarms introduce multiple advantages, they also raise different challenges that should be addressed. These challenges are related to the network deployment, coordination, and communication management, which become prohibitively complex as the network size increases. To perform more efficiently, the UAV surveillance system needs to manage two phases: The first one is the deployment of UAVs in the target region to monitor the incoming incidents \cite{pub.1101897582}. Each device controls a sub-region and interacts with others to ensure good coverage. Because of the low frequency of incidents' occurrences in some sub-regions, it is not worthy deploying UAVs in these zones while others located in hot sub-regions encounter high-frequency incidents \cite{LI201923}. The aim of the first phase is to maximize the coverage using a minimum number of devices \cite{DBLP:journals/ieeejas/HuangHM20}. The second phase consists of managing the coordination between UAVs either to collaboratively deal with incoming incidents or for path planning to ensure faster interventions \cite{8984371}. This phase will be widely investigated in this paper, specifically for surveillance scenarios. 

In surveillance applications, the aim is to monitor specific ground objects or identify threats within the target region. Thus, unlike satellite imagery, higher data resolution should be gathered, specially for a security plan \cite{8441965}. In this context, UAVs are the most suitable technology to provide such fine-grained data about the target object from different angles, which makes the identification more accurate \cite{8342149}. These data-generating devices are only responsible for collecting the data, while servers with higher capacities generate the identification results using Artificial Intelligence (AI) techniques. The traditional wisdom resorts to cloud or edge servers to compute heavy tasks. However, due to the harsh environments where UAVs flow (e.g., military border zones, forests, offshore oil reserves, etc.), the communication with remote servers is strongly affected by the weather. Also, the processing might be difficult or even impossible because of the interference resulting from the UAV altitude and the underground environment (e.g., high-rise building effect on path loss) \cite{Wang2019}. Furthermore, as UAVs are sending high-resolution images to cloud/edge servers at each small interval of time and knowing that incidents are rarely occurring, the  large  data  volume transmitted by source units has  become  problematic, particularly  for  systems  that  do not  have  stable  bandwidth  availability \cite{ 8766641}. Because of this tremendous amount of data obtained during the UAVs mission including those used to detect objects of interest or to perform accurate navigation, AI should be integrated into the design of UAVs \cite{Athanasis2018BigDA}.
%, 8939564, Opromolla2019AirborneVD

Deep Neural Networks used in computer vision to process data gathered by UAVs have been significantly improved in the last few years \cite{10.1016/j.eswa.2017.09.033}. This advancement has not been followed by the improvement of computation capabilities of UAVs. As an example, recent models are progressing toward deeper neural networks with higher computation demands, which is illustrated in table \ref{table:comparison}. Even though deep learning has achieved substantial breakthroughs in UAVs applications, it is still challenging to implement the complex DNN models in a single resource-constrained device, because of their computation requirements resulting in unfeasible execution time \cite{8894381}. 
\begin{center}
\begin{table}[!h]
\caption{Progression towards deeper neural network structures in recent years.}
\begin{tabular}{|l|l|l|l|l|}
\hline
\textbf{Architecture Name} & \textbf{Year} & \textbf{Parameters} & \textbf{Depth} & \textbf{Reference} \\ \hline
\textbf{LeNet}  & 1998 & 0.060 M & 5 & \cite{726791} \\ \hline
\textbf{AlexNet} & 2012 & 60 M & 8 & \cite{10.5555/2999134.2999257} \\ \hline
\textbf{VGG} & 2014 & 138 M & 19 &  \cite{simonyan2014deep}\\ \hline
\end{tabular}
\label{table:comparison}
\end{table}
\end{center}
\vspace{-1cm}

In order to fit the requirements of DNN into the resource-constrained devices, collaborative deep learning strategies have been recently proposed in the literature. The basic idea is to divide the DNN model into segments (e.g. layers, multiplication tasks, etc.) and each segment is allocated within a participant. Each participant shares the output to the next one until generating the final prediction. In this way, the entire inference can be locally performed at the proximity of the data source, without the need to transmit the original data to the remote servers. Nevertheless, the existing efforts mainly explored different possible partitioning that allows deploying DNN models on resource-limited devices (e.g., IoT devices, sensors, etc.). However, scheduling the communication between participants and designing a collaboration strategy to conduct inference tasks, while being constrained by computation and memory were not considered by previous works. Furthermore, accomplishing in-site classification by moving devices exposed to path loss and potential disconnection, has not been studied in the literature. Therefore, to support deep neural processing in UAV systems, the DNN distribution must be redesigned in order to consider hardware and physical constraints as well as planned paths and communications loss, which will be done in our work. 
%\cite{8493499}

In this paper, we study surveillance systems and we consider Convolutional Neural Networks (CNNs), as they demonstrated unprecedented efficiency for image classification. To match resource-consuming DNN solutions with the constraints of UAV devices, we leverage the hierarchical design characterizing deep learning models, in order to suitably place layers of different classification requests. Particularly, we propose an approach that receives as an input the set of CNN classification requests (whose model is previously trained), the technological characteristics of the deployed UAVs, and the planned paths for different devices; and gives as an output the optimal placement of layers within participants, while having as an objective to minimize the latency of classification tasks which is the delay spent for executing the distributed inference and the delay to communicate the intermediate data. To the best of our knowledge, we believe that we are the first to exploit the distributed resources within surveilling UAVs to perform local inferences without compromising the accuracy of results. Our proposed methodology is validated for multiple state of the art CNN models and different requests' loads and area sizes. The proposed approach also covers a large variety of applications, including public safety and control in smart cities where the transmission frequency is high or surveillance in harsh environment such as military zones and forests where the frequency bands are relatively low. The contributions of our paper are presented as follows:
\begin{itemize}
    \item We present our system model that consists of multiple drones, responsible for capturing scenes and processing CNN classifications.
    \item We formulate our UAV collaborative inference as a non-linear optimization problem that aims to minimize the classification latency with stationary participants (i.e. relative distances between UAVs are stationary), while respecting the limited available resources. Due to the complexity of the problem, we convert the non-linear integer optimization into a linear one following the big-M rules \cite{bigM}.
    \item To cover the non-homogeneous variation of the relative distances between UAVs, the stationary solution must be run multiple times to handle the topology variation of devices during the swarm movement, which introduces an additional delay to the decision-making latency and increases the complexity of the problem. For these reasons, we tailor this method to include the non-homogeneous mobility model in order to predict the future location of each UAV.
    \item We conduct extensive simulations to evaluate the
    performance of our approach under different network configurations, e.g. requests load, area dimension, etc. We illustrate that executing multiple requests using the moving devices is possible and we unveil different parameters that should be present to achieve the maximum performance, including the number of devices, their capacities, and the deployed CNN networks.
\end{itemize}
Our paper is organized as follows: Section \ref{related_works} explores relevant
related works. Section \ref{sec:formulation} presents our proposed framework and the problem formulation under stationary and dynamic environments. The experimental evaluation is provided
in section \ref{sec:results}. Finally, in section \ref{conclusion}, we draw the conclusions.
\begin{figure*}[h]
\centering
  \includegraphics[scale=0.7]{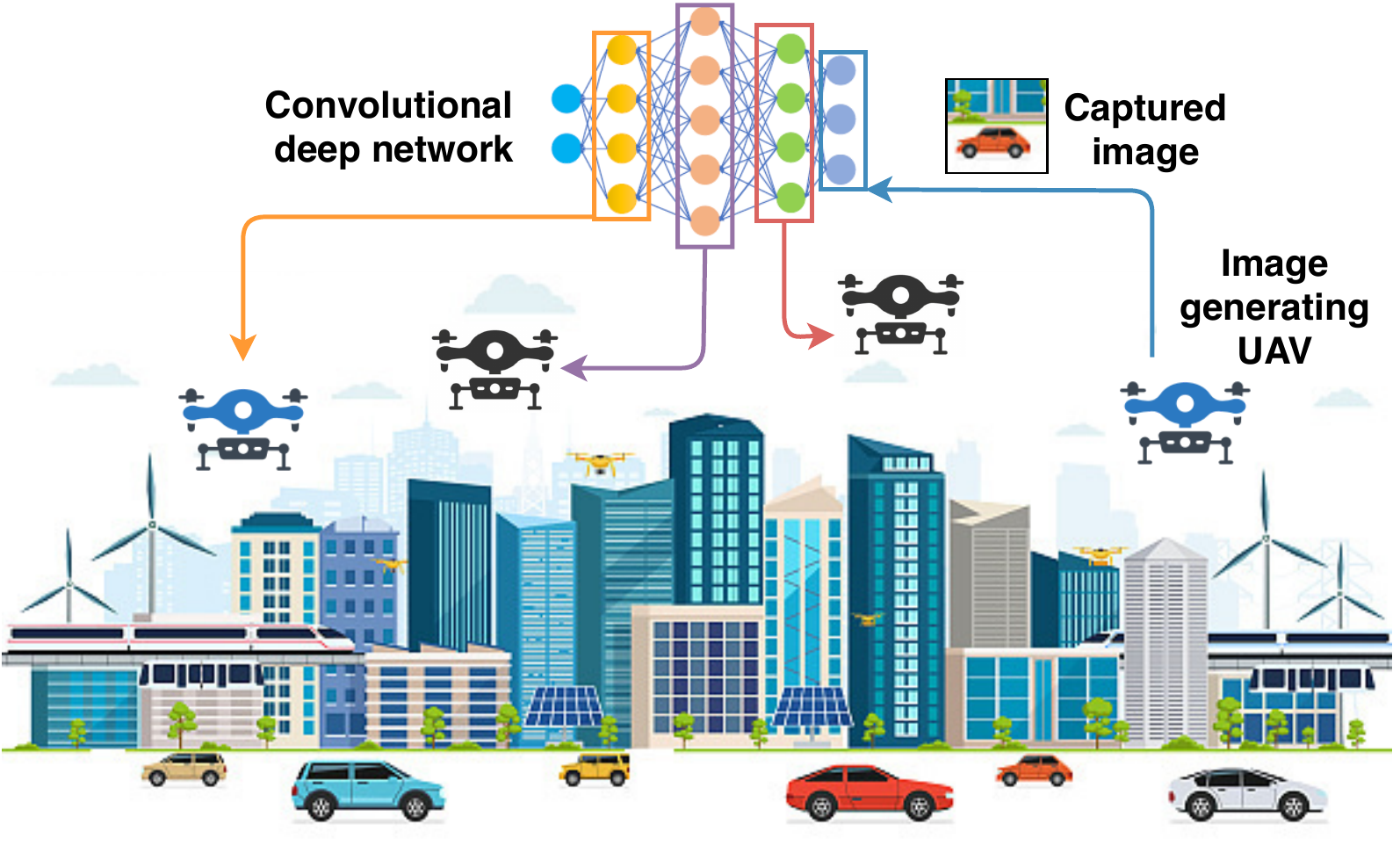}
  \caption{Illustration of a distributed  CNN  inference on resource-constrained  UAVs in a smart city surveillance system.}
  \label{system_model}
\end{figure*}

%========================================================
%========================================================
\section{Related Works}\label{related_works}
The advances in sensors' capabilities, e.g. cameras, endowed with the UAVs have enabled these devices to play a vital role in a new breed of services and applications that require unmanned operations. Recently, the integration of deep learning in UAV systems has leveraged the intelligence for addressing different challenging issues in these applications, and promising unprecedented performance and complexity reduction. As an example, authors in \cite{8421338} introduced a strategy to detect forest fires, relying on machine learning techniques. The proposed approach combines the high-performance resources of cloud servers, the rich resources of edge computing and the sensing capacities of drones to capture the incident scene, interpret it in the remote servers, and perform early forest fire detection. However, as UAVs are sending high-resolution images to cloud/edge servers in small intervals of time, the  large  data  volume transmitted by source units has  become  problematic, particularly for systems with fluctuating  bandwidth  stability. Because of these tremendous data obtained during the UAVs mission, several efforts proposed integrating AI into the design of UAVs, including the works in \cite{8756696} and \cite{8850815}. These efforts propose to utilize the on-board processing capabilities equipped with the UAVs to execute small-scale CNN networks such as YOLOv3-tiny. However, the testing results showed that the recognition accuracy is not sufficiently high. Additionally, due to the limited available resources, the classification latency  does  not  match  the  real-time  requirements  of some applications, particularly those that need prompt interventions.

Several efforts have been conducted to fit deep neural networks in resource-constrained devices, from storage, computation, or hardware perspective. A promising technique that has been proposed in the literature, is compression. Such an approach aims to compact the existing state of art DNNs, while minimizing the accuracy loss. Pruning \cite{8736011} is one of the compression approaches that consists of removing redundant weights/channels of smaller importance in the classification process. Quantization \cite{han2015deep} is another technique that proposes to reduce the number of bits required to represent the activation of the model. Finally, Binarization \cite{DBLP:journals/pr/QinGLBSS20} proposes to use binary values for weight and activation design. Even though the compression proved its capacity to integrate deep learning tasks in resource-constrained devices, this technique suffers from accuracy loss and cannot be applied to all types of dataset, neither supported by all devices. 

In order to fit the requirements of DNNs into small devices without sacrificing the accuracy of the results, collaborative deep learning strategies have been recently proposed. More specifically, the deep neural network is divided into segments and each segment is executed within a helper. Different IoT frameworks have been proposed for CNN partitioning such as TensorFlow \cite{abadi2016tensorflow}, DeepX \cite{7460664}, and Distributed Artificial Neural Networks for the Internet of Things (DIANNE) \cite{10.1145/2836127.2836130}. These solutions provide per layer distribution and require manual configuration from the user. In addition to the proposed distribution tools, multiple efforts have focused on the partitioning strategy that enables the implementation of large networks on small devices, e.g. sensors. Efforts dealing with DNN partitioning can be classified into three categories: First, many recent works, such as \cite{dis1} aimed at minimizing the transmitted data to the remote servers. Thus, they proposed to compute shallow layers on constrained devices and delegate deep layers to the cloud/edge computing. The logic behind this technique is that intermediate data generated by deep layers are reduced compared to the original data. Furthermore, shallow layers do not require high computation capacities. Second, other approaches adopted the hierarchical architecture, where the model is distributed among cloud, edge and mobile units \cite{7979979}. Even if resorting to remote computations can be a solution to minimize the delivered data, the transmission latency can prevent the implementation of such approaches, particularly in applications with low-delays requirements or with high classification demands. The third approach consists of relying only on pervasive devices to execute the  inference requests at the proximity of the data source. In this context, two partition techniques are introduced: either to adopt per-layer distribution or to apply per-input partitioning (e.g., rows or columns of feature maps). Authors in \cite{disabato2019distributed} proposed per layer CNN distribution  for IoT systems, where devices are known by their technological constraints. These resource-limitations prevent them from locally executing the inference on the gathered data. Hence, a methodology to distribute multiple CNNs is proposed while optimizing the latency between data gathering and decision making. 
%. Although, the waiting time and the remote connectivity breaks might result in the infeasible delay of decision making. The IoT devices in this work are distinguished by data sources and computing units such as a 
Authors in \cite{7927211} proposed a per-input partition scheme for pre-trained DNN models to fit with resource-constrained mobile devices and accelerate DNN computations. In this method, the data delivery time between devices is minimized while considering two types of DNN layers and heterogeneous mobile devices  which are used as computing resources. %Otherwise, the main drawback is that only one device can act as a data source and designed to be the group owner. 
A study conducted in \cite{6706764} proved that wireless sensor networks can be used as a hardware platform to implement parallel and distributed neurocomputing. This architecture is designed by mimicking the biological neural networks that exist in the brain. In this way, each sensor serves as the processing unit of an artificial neuron. DeepThings framework proposed in \cite{8493499} provides adaptively distributed inference of CNN model on tightly resource-constrained IoT edge clusters. This framework uses the Fused Tile Partitioning (FTP), which is an input-wise division technique designed to reduce the memory footprint while ensuring parallelism. 

Our proposed method is distinguished from the aforementioned approaches by several points: (1) The efforts that focused on re-designing the model or exploiting the hardware efficiently to fit the CNN networks in resource-constrained devices, such as compression, are sacrificing the accuracy. This is not the case of our approach that preserves accuracy, as no modification of the CNN characteristics is introduced during the distribution. Instead, we distribute a high-performance trained model with proven efficiency for scene classification. %while the intermediate data exchange is ensured through wireless communication (only connected UAVs for a time duration are involved). 
(2) The input-wise distribution or the per-neuron distribution are designed for highly limited-resources IoT devices or for sensor networks that are not even able to execute layers of advanced CNN models. Furthermore, these methods are implemented in large pervasive systems that include hundreds of devices and  require a large dependency between participants to receive the outputs of different segments. This strategy cannot be applied to the UAV networks due to their low density (small number of drones) and the cost of dependency on air-to-air communication in terms of connectivity, latency, and consumed bandwidth. (3) The previously-described works proposed a static distribution of CNN models, such as each subset of devices is responsible to process a specific portion of the CNN model. This does not fit our online system, where a dynamic incoming load is received randomly during the surveillance mission and the connected UAVs must cooperate to autonomously parallelize all requests and optimally leverage the available resources, without being constrained by a static planned partitioning. Moreover, UAVs are moving units that can encounter disconnection situations, which makes static distribution invalid in our scenario. (4) To the best of our knowledge, we are the first to investigate real-time CNN inference distribution over a UAVs network, while considering the distinctive characteristics of air-to-air communication.

%==================================================================
%==================================================================
\section{Distributed CNN on  resource-constrained UAVs}\label{sec:formulation}
In this section, we present our distributed convolutional
neural network approach for resource-constrained UAV devices and low decision-latency applications. For that, we start by designing our system model. Subsequently, we formulate this problem as an optimization, aiming at minimizing the total latency to execute the incoming classification requests. As we are dealing with moving devices, we next update our model to consider a UAVs' mobility model and the data-rate variation of different participants.

\subsection{System model}
We consider a group of UAVs that are capable of forming a wireless ad hoc network. The network is composed of a set of $N$ UAVs, namely $\mathbb{N}_N$, as illustrated in Figure \ref{system_model}. These UAVs are responsible for monitoring a target area by capturing images and requesting data classification (e.g. detecting forest fires, detecting highway accidents, etc.). Two types of communication may occur in our network: UAV to UAV (U2U) and UAV to Ground device (U2G or Ground device to UAV) in order to connect to remote servers and send incident alarms. In our work, we will focus only on the U2U communications. The UAVs, deployed in the region under surveillance, are equipped with different IoT devices, such as cameras and GPS \cite{pham2018autonomous}. Moreover, each one of them can be either a data-generating node or/and a computation unit. Particularly, the data-generating node is responsible for gathering the data using its embedded camera and requesting the inference of the captured image, while the computation unit is a UAV performing a sub-task of the inference. Each UAV $i \in \mathbb{N}_N$ is characterized by a limited memory $\bar{m_i}$ and computation capacity $\bar{c_i}$, preventing it from performing the inference independently onboard. Such resource-constrained UAVs may delegate some subsequent inference tasks to the neighboring UAVs, in order to perform a complete classification locally.

We denote the 3D coordinates of each UAV $i \in \mathbb{N}_N$ by ($x_i,y_i,h_i$) and we assume for simplicity that all UAVs fly at a fixed altitude above the ground, namely $h_i=H$, while considering the minimum $H$ value required for safety (e.g., building or structure avoidance) \cite{8531711}. The horizontal coordinates ($x_i,y_i$) of a UAV $i$ vary over time following the trajectory of the master node. In this way, each UAV deployed in the network moves from its initial position to the final destination in a cyclic trajectory to cover the target area. The positions of UAVs are recorded periodically at each $T$ time step. This means that the position of each UAV is repeated every $T$ seconds.
Therefore, the optimization introduced in the next section is executed periodically to handle the network variation (connectivity and U2U link quality) and the dynamic incoming requests.

In this work, we consider a swarm of UAVs cooperating to execute a single CNN model, which means we are dealing with homogeneous input data acquired by these UAVs and requiring classification. Let $M$ define the number of layers characterizing the considered CNN model, in other terms, it defines the number of sub-tasks that should be distributed among the available UAVs and executed to perform the classification of the input data. Each layer of the CNN model is characterized by a memory requirement and a computation demand $c_i$. In this context, different UAVs are endowed with a copy of the trained model, allowing them to execute any assigned sub-task. Let $K_j$, $\forall \quad j \in \{1,..., M\}$, be the size of the intermediate output generated by layer $j$ and transmitted to the subsequent layer $j+1$ of the CNN model, and let $K_s$ denote the memory occupation of the input image acquired by the source node and requiring classification. This image is potentially transmitted to a neighboring UAV for the execution of the first layer. Furthermore, each UAV $i \in \{1,..., N\}$ generates a number of requests $R_i$, such as $0\leq R_i \leq R$.

As the G2U links are only used to offload the classification results in our scenario, our work will focus only on scheduling the U2U communication between different UAV participants to perform classification tasks. Specifically, a U2U link is established between any two UAVs $i$ and $k\in \mathbb{N}_N$, which are characterized by the estimated achievable data rate defined as follows \cite{8654727}:
\begin{equation}
\label{equ:rho}
    \rho_{i,k}= B_i.\log_2(1+\Gamma_{i,k}),
\end{equation}
where $B_i$ denotes the bandwidth of the UAV device $i$ and $\Gamma_{i,k}$ is the average Signal-to-Interference-plus-Noise Ratio (SINR) of the U2U link between UAVs $i,k\in \mathbb{N}_N$.

Next, we will design an optimal placement of different
layers in the UAV units participating in our distributed
and collaborative system, while considering
the resource capabilities of the devices. As previously described, the optimization is executed periodically to cover the variation of the network density and the UAVs location. We emphasize that  the proposed distribution encompasses only standard CNN models with known types of processing (e.g., convolutional, ReLU, etc.), and without any residual block \cite{eshratifar2019collaborative}.

\subsection{Optimal UAV-based layer distribution (OULD)}\label{sec:method}

In this section, we introduce the proposed methodology for the optimal placement of different classification sub-tasks incoming from data-generating UAV nodes, with an objective to minimize the total latency in decision making while considering only onboard UAVs. This latency is defined as the required time to transfer the output of layers between participants and to compute different tasks. We formally define the distribution problem as an optimization problem subject to the UAVs computation constraints. Our objective in this paper is to minimize the latency of classification tasks, which is the delay spent for executing different layers and the delay to communicate their intermediate data. The proposed strategy relies on 3D matrix of decision variables, namely $\alpha_{r,i,j}$, that returns 1 if the UAV $i$ executes the layer $j$ of the request $r$, and 0 otherwise:
\begin{equation}\label{equ:alpha}
    \alpha_{r,i,j} =
  \begin{cases}
    1  & \quad \text{if UAV } i \text{ executes the layer } j \text{ of request } r.\\
    0  & \quad \text{otherwise.}
  \end{cases}
\end{equation} 

The objective function models the end-to-end latency spent to classify all incoming requests R to the network composed of N UAVs using collaborative inference. Such as the execution of each request is composed of M subtask depending on the type of incoming request, knowing that a CNN model is composed of M layers. The end-to-end latency (or decision-making latency) is defined as the time between gathering images of size $K_s$ by the source nodes and the decision-making. The objective function to be minimized is formulated as follows:
% Hence, the proposed solution relies on the $RNM$ variables,
\begin{equation}\label{eq:objective}
   \underset{ \begin{subarray}{c}
    (\alpha_{r,i,j}) 
    \end{subarray}}\min \sum_{r=1}^{R}\sum_{i=1}^{N}\sum_{\substack{
   k=1 \\
   k\neq i}}^{N}\sum_{j=1}^{M-1} \alpha_{r,i,j}.\alpha_{r,k,j+1}.\frac{K_j}{\rho_{i,k}} + t_s
\end{equation}

s.t

\begin{equation}\label{equ:layers}
    \forall i \in \mathbb{N}_N \quad \quad  \sum_{r=1}^{R}\sum_{j=1}^{M} \alpha_{r,i,j}.m_j \leq \bar{m_i}
\end{equation}

\begin{equation}\label{equ:memory}
    \forall i \in \mathbb{N}_N \quad \quad  \sum_{r=1}^{R}\sum_{j=1}^{M} \alpha_{r,i,j}.c_j \leq \bar{c_i}
\end{equation} 

\begin{equation}\label{equ:once}
    \forall r \in \mathbb{N}_R, \forall j \in \mathbb{N}_M \quad \quad  \sum_{i=1}^{N} \alpha_{r,i,j} = 1
\end{equation}

\begin{equation}\label{equ:binary}
    \forall r \in \mathbb{N}_R, \forall i,j \in \mathbb{N}_M \quad \quad   \alpha_{r,i,j} \in \{0,1\}
\end{equation}
where 

\begin{equation}\label{equ:ts}
    t_s = \sum_{r=1}^{R}\sum_{i=1}^{N}\sum_{\substack{
   k=1 \\
   k\neq i}}^{N} \mu_{i,r}.\bar{\alpha}_{r,i,1}.\alpha_{r,k,1}.\frac{K_s}{\rho_{i,k}}
\end{equation}

In the objective function (\ref{eq:objective}), we define two different components of the latency: 

(1) The source latency $t_s$ defined in Eq. (\ref{equ:ts}), which is the time required by the UAV source to transmit its gathered image to the UAV executing the first layer of this request. We remind that each UAV has a number of requests $R_i \in \mathbb{N}_R = \{0,...,R\}$, such that $R=\sum_{i=1}^{n}R_i$, and $\bar{\alpha}_{r,i,1}$ is the complement of the decision variable $\alpha_{r,i,1}$.

(2) Time spent on transmission of the intermediate outputs between UAVs responsible of different tasks. More formally, the transmission time of the intermediate output of layer $j$ between UAVs $i$ to $k$ is given by $ \frac{K_j}{\rho_{i,k}}$, where $\rho_{i,k}$ is the data rate of the U2U link between UAVs $i$ and $k$ and $K_j$ is the size of the generated output from layer $j$ of the CNN model. 

The wireless communication is the key characteristic that distinguishes a UAV system from IoT or  terrestrial ad-hoc networks, where the machine learning distribution is widely studied. The air-to-air communication channel is known by the high probability of line of sight that  depends on the altitude of UAVs, which is not the case of terrestrial networks.

The constraint in Eq. (\ref{equ:layers}) ensures that each UAV executes multiple layers from multiple requests, while not exceeding its memory limit. Eq. (\ref{equ:memory}) guarantees that the constraint on computational load is respected for all  participants, while assigning the classification tasks. The constraint in Eq. (\ref{equ:once}) implies that each layer $j \in \mathbb{N}_M$ for each request $r \in \mathbb{N}_R$ is executed once in a single UAV. This constraint is designed to avoid redundant tasks execution. Finally, the constraint in (\ref{equ:binary}) ensures that the decision variable is binary and takes only a value of $0$ or $1$.

We can see that the objective function in the optimization problem (\ref{eq:objective}) is nonlinear (NLP) and non-convex. More specifically, a problem that involves non-convex objective function/constraints may have multiple feasible regions and local optimal points, which makes solving it extremely complex and it may take exponential time in the number of variables and constraints to determine the global optimum across all regions. Therefore, in our work, we try to ensure the convexity and linearity of the optimization. More specifically, we introduce a new decision variable, namely $\gamma_{r,i,k,j}$, defined by:

\begin{equation}\label{equ:linear}
\begin{aligned}
    \forall r \in \mathbb{N}_R, \forall i,k \in \mathbb{N}_N, \forall j \in \mathbb{N}_M \\  \gamma_{r,i,k,j}=\alpha_{r,i,j}.\alpha_{r,k,j+1},
\end{aligned}
\end{equation}
where $\gamma_{r,i,k,j} \in \{0,1\}$ is the big M rule parameter used to convert the non-linear integer problem to linear, which is defined in our case as the transmission of the output of the layer $j$ computed in the UAV $i$ to the next participant $k$. Particularly:
\begin{equation}\label{equ:gamma}
   \gamma_{r,i,k,j} =
  \begin{cases}
    1  & \quad \text{if UAV } i \text{ executes the layer } j \text{ of request }  \\
    &\quad r \text{ and sends the output to the UAV } k \text { to}\\
     &\quad \text{compute the next layer } j+1.\\
    0  & \quad \text{otherwise.}
  \end{cases}
\end{equation}
To ensure that $\gamma_{r,i,k,j}$ is equal to $\alpha_{r,i,j}.\alpha_{r,k,j+1}$, three constraints should be considered following the big M integer and linear programming rule \cite{bigM}:
\begin{equation}\label{equ:BigM}
\begin{aligned}
    \forall r \in \mathbb{N}_R, \forall i,k \in \mathbb{N}_N, \forall j \in \mathbb{N}_M \\  \gamma_{r,i,k,j} \leq \alpha_{r,i,j},\\
    \gamma_{r,i,k,j} \leq \alpha_{r,k,j+1},\\
    \gamma_{r,i,k,j} \geq \alpha_{r,i,j}+\alpha_{r,k,j+1}-1.
\end{aligned}
\end{equation}
These constraints ensure that $\gamma_{r,i,k,j}$ is equal to $0$, if $\alpha_{r,i,j}$ or $\alpha_{r,k,j+1}$ is null and equal to 1 if both of them are equal to 1. In this way, our new Integer Linear Programming (ILP) optimization will include two decision variables  $\gamma_{r,i,k,j}$ and $\alpha_{r,i,j}$ and the objective will be reformulated to:
\begin{equation}\label{eq:objectiveLinear}
   \underset{ \begin{subarray}{c}
    (\alpha_{r,i,j},\gamma_{r,i,k,j}) 
    \end{subarray}}\min \sum_{r=1}^{R}\sum_{i=1}^{N}\sum_{\substack{
   k=1 \\
   k\neq i}}^{N}\sum_{j=1}^{M-1} \gamma_{r,i,k,j}.\frac{K_j}{\rho_{i,k}}  + t_s
\end{equation}

where 

\begin{equation}\label{equ:ts+}
    t_s = \sum_{r=1}^{R}\sum_{i=1}^{N}\sum_{\substack{
   k=1 \\
   k\neq i}}^{N} \mu_{i,r}.\gamma_{r,i,k,1}.\frac{K_s}{\rho_{i,k}}
\end{equation}
Following the changes introduced to make our problem linear and convex, finding the optimal solution becomes less complex and less time consuming.
\begin{figure}[t]
\centering
\begin{subfigure}[c]{1.0\columnwidth}
		\centering
		\includegraphics[width=\linewidth]{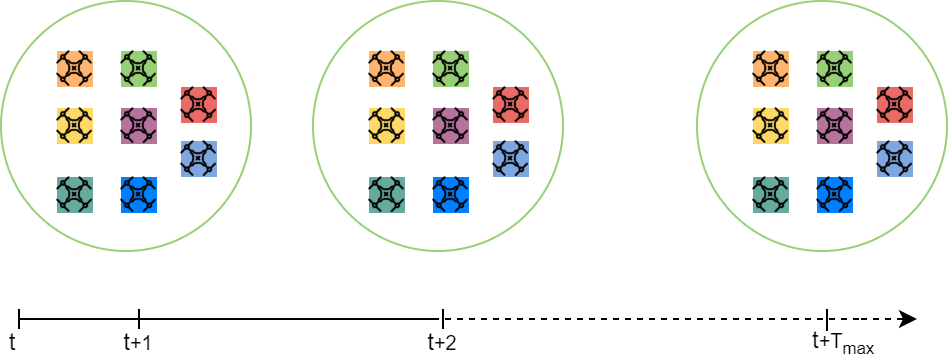}
		\caption{Homogeneous mobility}
		\label{fig:Homogeneous}
\end{subfigure}\\
\begin{subfigure}[c]{1.0\columnwidth}
		\centering
		\includegraphics[width=\linewidth]{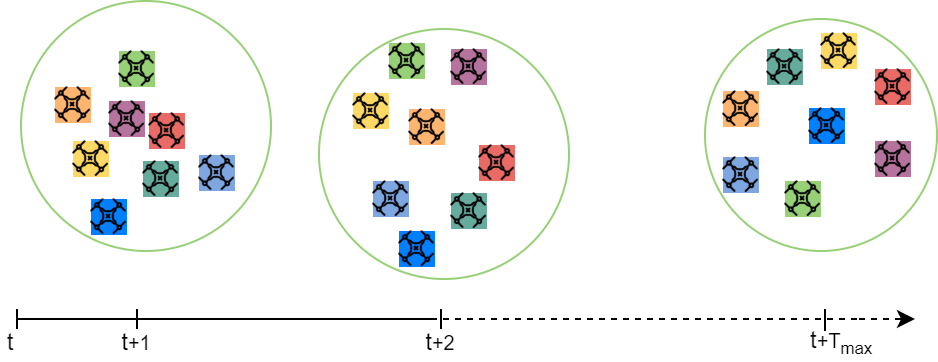}
		\caption{Non-Homogeneous mobility}
		\label{fig:non-Homogeneous}
\end{subfigure}
\caption{A scenario of UAVs moving among the target region.}
\label{fig:mobility}
\end{figure}

\subsection{OULD with mobility prediction (OULD-MP)}\label{sec:tailoring}

The distribution approach described in the previous section gives efficient results for the homogeneous mobility model (Figure \ref{fig:Homogeneous}), where the relative distance between UAVs remains static during the swarm movement in the target region. Such as figure 2 shows the positions of all UAVs at each time step t for both scenarios homogeneous and non-homogeneous which are distinguished by the relative distance between the UAVs involved in the inference process. The use of OULD method in the non-homogeneous mobility scenario requires multiple runs for each network configuration while the swarm is moving through time. This increases the complexity of the optimization solution, gives a sub-optimal distribution policy, and introduces an additional delay in the decision-making latency. Thus, when the relative distance between UAVs changes, the OULD distribution policy does not present an efficient solution due to the change in data rate-based distance between UAVs. In this section, we tailor the proposed approach to cover the non-homogeneous mobility model (Figure \ref{fig:non-Homogeneous}) in a one shot optimization, by considering a mobility prediction for the time steps $t \in \{1..T\}$ which consists of adopting a mobility model to predict the future locations of UAVs at each time step $t$. This allows us to perform a single distribution by solving the optimization problem, while considering the future locations of UAVs. This results in an optimal solution that is efficient for, not only the current time step, but also for some future ones. This helps us in avoiding to run the optimization problem each time step (knowing its complexity and the execution time) to handle data-rate variation between UAVs that tremendously influences the performance of the inference task distribution in terms of end-to-end latency. The mobility model plays a key role in the design of cooperative applications involving swarm of UAVs, such as crowd or region monitoring as well as critical platform surveillance. More precisely, the mobility model relies on different constraints, such as the energy constraint for path planning, area coverage, and network connectivity. The latter constraint has a substantial impact on our system since the loss of an intermediate task assigned to a disconnected UAV due to communication outage implies the loss of the inference of the corresponding request.
In this paper, the distances, connectivity, and movements of UAVs, as well as the swarm participants (e.g. joining and leaving devices) that depict the variation of the network at each time step are modeled through the communication link quality. Indeed, the link quality will be represented by the transmission data rate $\rho_{i,k}$ which denotes the expected amount of data transmitted between 2 UAVs $i$ and $k$ per movement period. More specifically, as shown in equation \ref{equ:rho}, one of the main factors affecting the physical data rate $\rho_{i,k}$ of a wireless link is the SINR ratio, which depends on the path loss. Because of this path loss, the received signal power is proportional to $d_{i,k}^{-\alpha}$, where $d_{i,k}$ is
the distance between the transmitter $i$ and receiver $k$ and $\alpha$ is
the path loss exponent. In this way, when $d_{i,k}$ is higher, the related data rate is lower. Meaning, lower data rates correspond to distant UAVs and vis-versa. Moreover, if the data rate between two UAVs becomes larger over time, it means that these devices are approaching each other. Finally, if the distance is very high and causes wireless link disconnection, the path loss and SINR can be approximated to 0 and $\rho_{i,k}$ will be accordingly equal  to $B_i.log_2(1)=0$. To summarize, the data rates characterising different participants can be used to define the distances between them, the possibility to perform data transmission, and their positions and movements over time.

\begin{figure}[t]
\centering
\begin{subfigure}[c]{1.0\columnwidth}
		\centering
		\includegraphics[width=\linewidth]{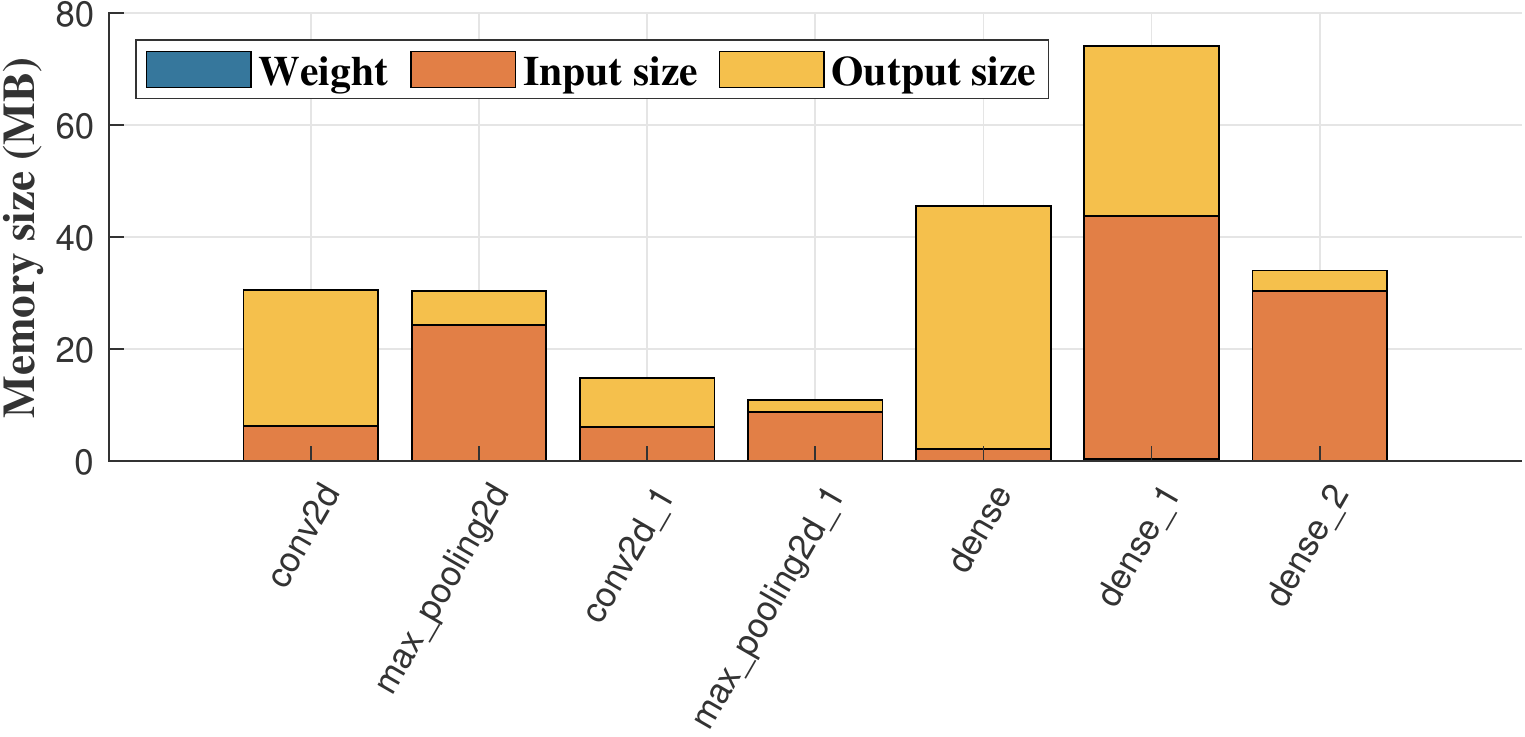}
		\caption{Lenet}
\end{subfigure}
%\hfil{}
\begin{subfigure}[c]{1.0\columnwidth}
		\centering
		\includegraphics[width=\linewidth]{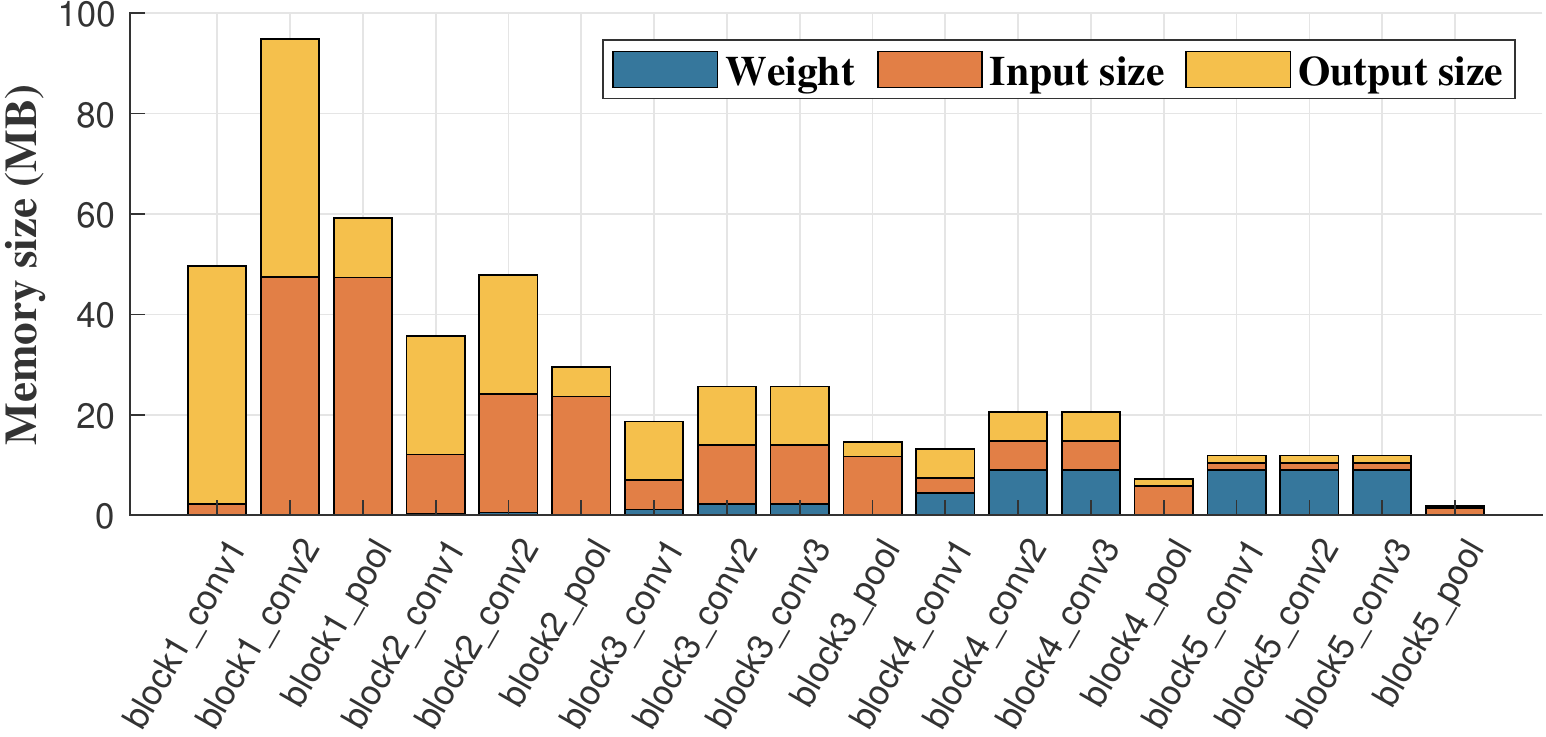}
		\caption{VGG-16}
\end{subfigure}
\caption{Inference memory footprint.}
\label{fig:inf-models}
\end{figure}

Various mobility models are suggested in the literature to match different application requirements. In our work, we will adopt the Reference Point Group (RPG) mobility model introduced in \cite{8207649}, which is the most suitable model for target region surveillance and fits our intention of distributing inference tasks among a set of UAVs. More specifically, this mobility model is designed to manage the behavior of the swarm that needs to follow a logical center or the group leader. All the UAVs in this group are randomly distributed around the reference node and combine their own mobility models with that of the group leader, which makes them follow the direction of the group,  with a small range of liberty. In the RPG model, the group leader follows a round trip movement between an initial and a final point, chosen in such a way to cover the entire target area. The movement of the leader UAV defines not only its motion but also the motion trend of all UAVs inside the group. Otherwise,  UAVs within the group can deviate from the planned path and lose the global motion trend, even if they are still in the group range. To enhance our static distribution methodology, we will consider the predicted locations of all UAVs in the network, at each time step for a time duration $T$. The same decision variables $\alpha_{r,i,j}$ and $\gamma_{r,i,k,j}$, defined in Eq.(\ref{equ:alpha}) and Eq. (\ref{equ:gamma}) respectively, will be used in this section, since the objective remains the same and the extension of the model resides only in finding an adaptive distribution of the CNN tasks between a moving set of UAVs for a certain time duration. In other words, $\alpha_{r,i,j}$ and $\gamma_{r,i,k,j}$ in this case present the inference task $j$ of the request $r$ assigned to UAV $i$ and the transmission of the $j$ layer output between two devices during a period of time $T$. Hence, the objective function defined in Eq.(\ref{eq:objectiveLinear}) is reformulated as follows:

\begin{equation}\label{eq:objective+}
   \underset{ \begin{subarray}{c}
    (\alpha_{r,i,j},\gamma_{r,i,k,j}) 
    \end{subarray}}\min \sum_{t=1}^{T}\sum_{r=1}^{R}\sum_{i=1}^{N}\sum_{\substack{
   k=1 \\
   k\neq i}}^{N}\sum_{j=1}^{M-1} \gamma_{r,i,k,j}.\frac{K_j}{\rho_{i,k}(t)}  + t_s
\end{equation}

where 

\begin{equation}\label{equ:ts+}
    t_s = \sum_{r=1}^{R}\sum_{i=1}^{N}\sum_{\substack{
   k=1 \\
   k\neq i}}^{N} \mu_{i,r}.\gamma_{r,i,k,1}.\frac{K_s}{\rho_{i,k}(t)}
\end{equation}

The problem constraints defined in Eq.(\ref{equ:memory}), (\ref{equ:layers}), and  (\ref{equ:once}) will not be reformulated, as the UAVs' mobility model during the mission does not impact the memory occupancy and the computation capability. These constraints depend only on the distribution map $\alpha_{r,i,j}$, the capacities of devices and the resource requirements of different layers. The objective function defined in Eq.(\ref{eq:objective+}) is composed of two latencies. The first part presents the time spent to transmit the intermediate outputs while considering the movement of UAVs during the time steps $t \in \{1..T\}$. This mobility results in a data rate variation while transmitting the layers outputs between any two UAVs $i$ and $k$, which is presented by $\rho_{i,k}(t)$. Thus, the variation of the relative distance between UAVs during swarm mobility influences the intra-communication quality, consequently impacts the end-to-end latency through intermediate outputs transmission.

\section{Performance evaluation}\label{sec:results}

In this section, we evaluate the performance of our distribution methods under different configurations of the UAV network (e.g. capabilities of the UAVs, trained CNN network, covered area, etc.). The proposed methodology has been simulated by solving the optimization problem defined in Eq. (\ref{eq:objective}) and (\ref{eq:objective+}). Due to the complexity of our combinatorial optimization solution, we used an High Performance Computer cluster to run the above simulation. We focused in our study on two main parameters 1) The average end-to-end delay, which is defined as the average latency per request required to classify one captured image by the collaborative swarm of UAVs; 2) The amount of data shared between all participants to accomplish all CNN sub-tasks.

Our system is composed of heterogeneous UAVs equipped with Raspberry Pi \cite{8472476} units that use their limited resources to participate in performing cooperative inference on the incoming classification requests. These resource-constrained UAVs are communicating with each other to ensure collaborative inference using the bandwidth $B_i$ = 20 MHz. As the CNN depth has a great impact on the number of requests that can be classified by leveraging the available resources, we are examining throughout this simulation the performance of our system on two CNN models; namely Lenet composed of 7 layers and VGG-16 \cite{simonyan2014deep} that comprises 18 layers. We adopt in our system a pedestrians surveillance scenario, where we classify 595x326 RGB images from the Stanford Drone Dataset \cite{StanData}. The memory requirements per layer of each CNN model to infer one image are shown in Figure \ref{fig:inf-models}. These footprints of different layers are calculated using Keras in Google collab.  The Figure shows that the memory demand of the CNN model including that of the input, intermediate, and output data as well as the size of millions of parameters (weight and biased) may prevent it from being executed on a single UAV.

\begin{figure*}[t]
\centering
\begin{subfigure}[c]{1.0\columnwidth}
		\centering
		\includegraphics[width=\linewidth]{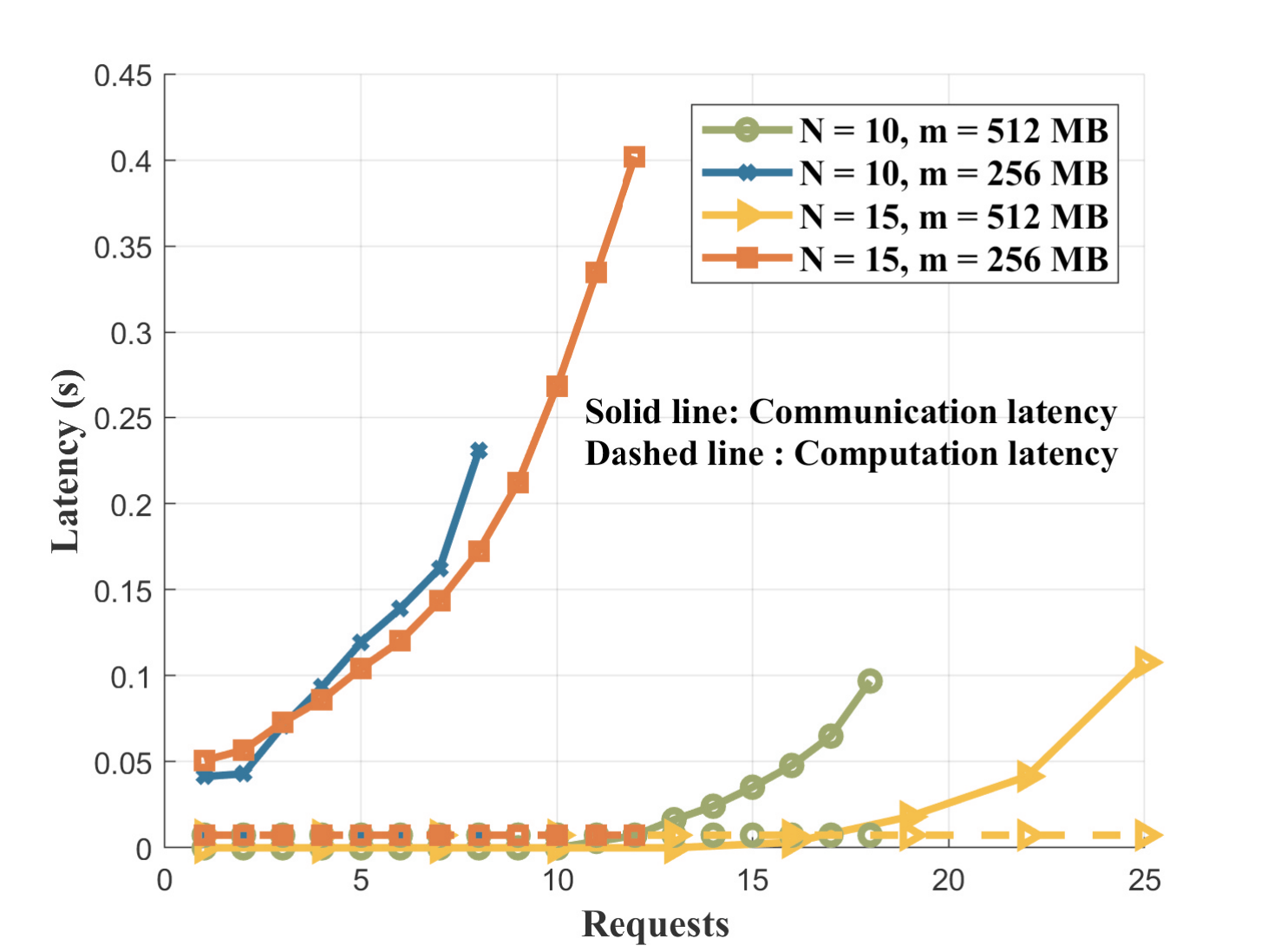}
		\caption{Average end-to-end latency per request.}
		\label{sub-fig:lenet_OULD_latency}
\end{subfigure}
\hfil{}
\begin{subfigure}[c]{1.0\columnwidth}
		\centering
		\includegraphics[scale=0.6]{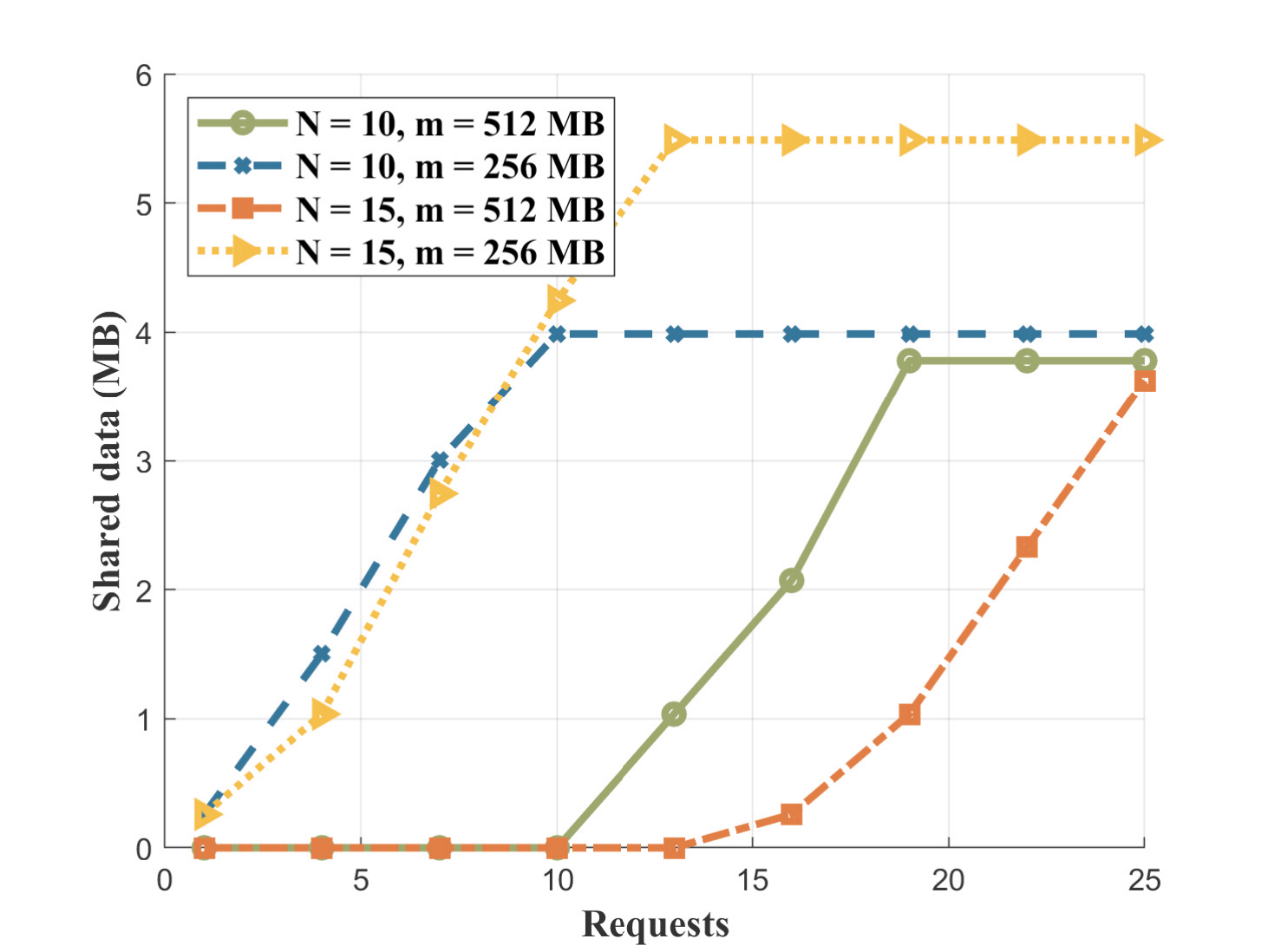}
        \caption{Shared data.}
		\label{sub-fig:lenet_OULD_sharedData}
\end{subfigure}
\caption{Performance evaluation of our proposed distribution method, namely OULD, while varying the network density N and the memory capability m. Two parameters are studied: the average latency per request and the shared data transmitted between the involved UAVs in the cooperative inference.}
\label{fig:lenet_OULD}
\end{figure*}

\subsection{OUD evaluation}
In this section, we conduct a distribution of classification requests on a homogeneous network mobility, where the UAVs' relative distance remain static over traveling time, as presented in section \ref{sec:method}. We emphasize that a key advantage of our approach is that it does not impact the accuracy of the model during the distribution process. Indeed, our strategy guarantees the connection between UAVs participating in the distribution, so that the CNN model parameters are not distorted as intermediate data losses are not allowed. In this simulation, we are considering multiple CNN models distribution for an online system where dynamic incoming load requires real-time inferences. We concentrate our efforts on exploring the impact of our proposed method on the average latency per request and the shared data to study the efficiency of the obtained distribution policy on the performance of UAVs-based surveillance systems. While there are many parameters of importance, these two are critical to highlight in our work due to these surveillance systems that require real-time inference and are evaluated in general based on the decision-making latency as well as the cost of this process which is presented in our work as the shared data exchanged between UAVs to accomplish the distributed inference.

Figure \ref{fig:lenet_OULD} illustrates the distribution of Lenet incoming requests over a network of 10 UAVs , where the impact of memory and computation constraints is examined. We evaluate the performance of our distribution strategy on different technological Raspberry Pi micro-computer families \cite{raspberrypi-models-comparison}, where we consider two levels of memory capabilities; a high-level memory equal to 512 MB and a low-level memory equal to 256 MB. Moreover,  due to the static impact of the computation capacity, we are considering one computation level equal to 9.5 GFLOPS.

\begin{figure*}[!ht]
     \begin{minipage}[l]{1.0\columnwidth}
\centering
	\begin{subfigure}[b]{\linewidth}
		\centering
		\includegraphics[width=\textwidth]{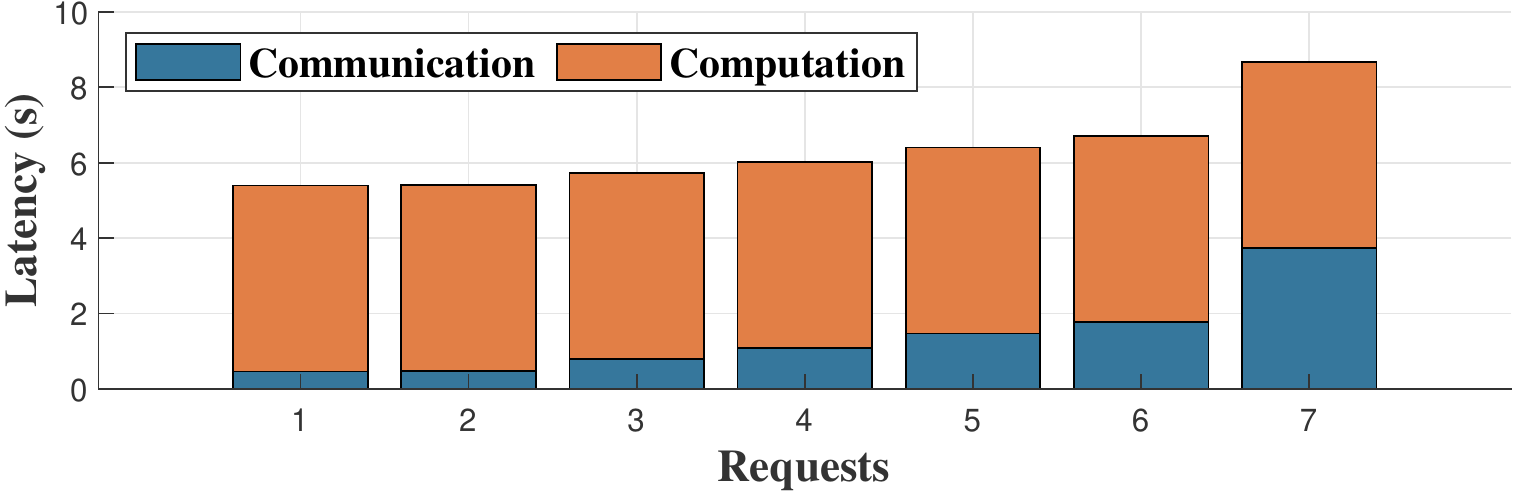}
		\caption{m = 512 MB and c = 9.5 GF}
		\label{sub-fig:vgg-10-HH}
	\end{subfigure}\hfill
	\begin{subfigure}[b]{\linewidth}
		\centering
		\includegraphics[width=\textwidth]{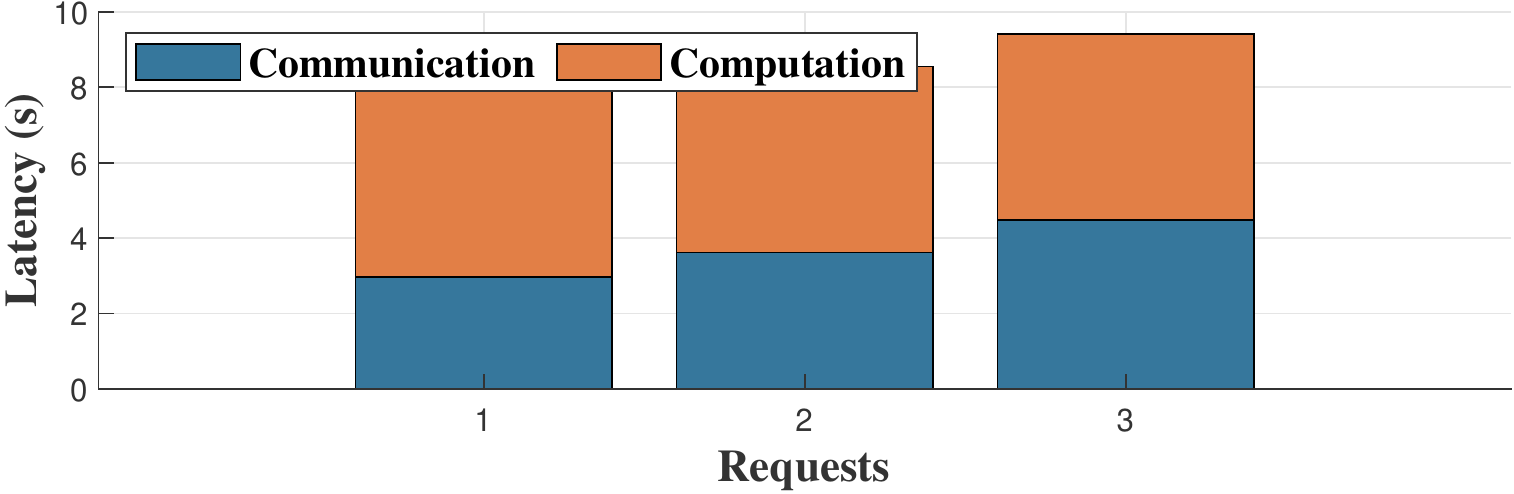}
		\caption{m = 256 MB and c = 9.5 GF}
		\label{sub-fig:vgg-10-LH}
	\end{subfigure}\hfill
	\caption{Optimal decision-making latency of a distributed VGG's inference over a network of 10 UAVs.}
	\label{fig:vgg-10}
     \end{minipage}
     \hfill{}
     \begin{minipage}[r]{1.0\columnwidth}
         \centering
	\begin{subfigure}[b]{\linewidth}
		\centering
		\includegraphics[width=\textwidth]{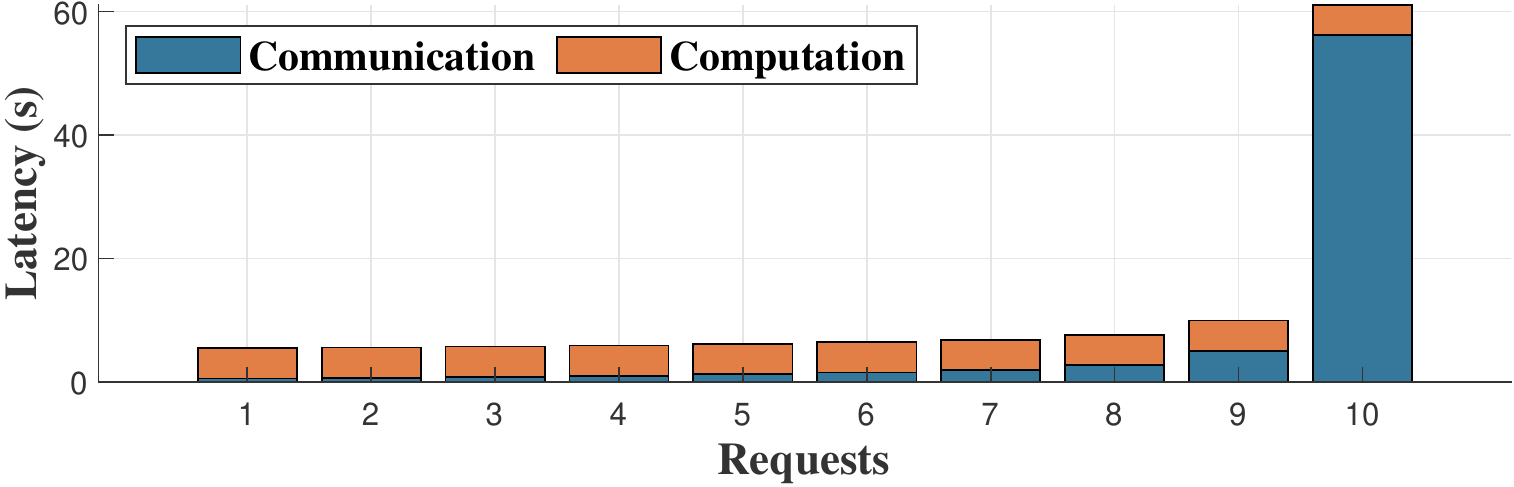}
		\caption{m = 512 MB and c = 9.5 GF}
		\label{sub-fig:vgg-15-HH}
	\end{subfigure}\hfill
	\begin{subfigure}[b]{\linewidth}
		\centering
		\includegraphics[width=\textwidth]{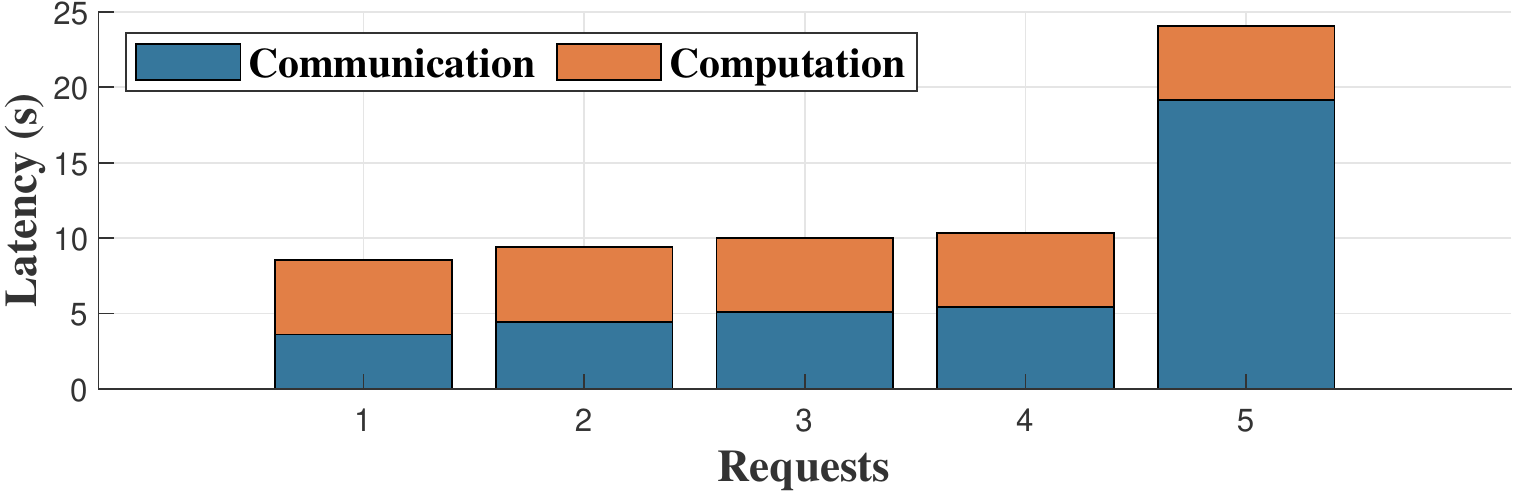}
		\caption{m = 256 MB and c = 9.5 GF}
		\label{sub-fig:vgg-15-LH}
	\end{subfigure}\hfill
	\caption{Optimal decision-making latency of a distributed VGG's inference over a network of 15 UAVs.}
	\label{fig:vgg-15}
     \end{minipage}
\end{figure*}

\begin{figure}[t]
\begin{center}
\includegraphics[scale=0.6]{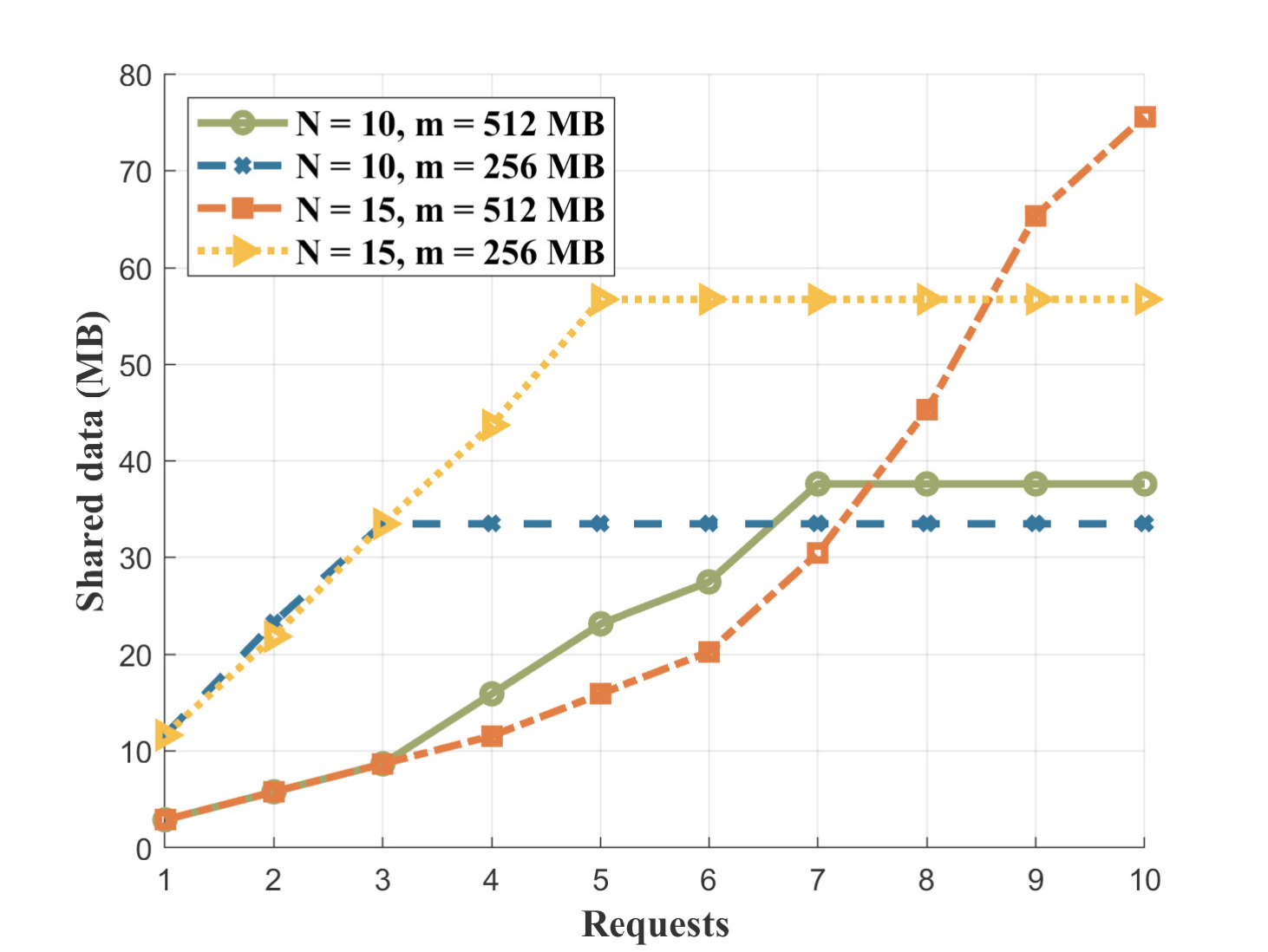}
\caption{Shared data of VGG-16 distribution.}
\label{fig:vgg-data}
\end{center}
\end{figure}

The solid green line in Figure \ref{sub-fig:lenet_OULD_latency} illustrates the optimal communication latency per request of a system that comprises 10 UAVs with high memory, deployed within a target area dimension equal to 100x100 m. The plot shape shows that this swarm of devices can cooperate to parallelize the classification of up to 18 requests. However, when the number of online incoming requests is high, the latency increases due to the increase of the intermediate exchanged data. In fact, the communication delay to transmit intermediate data generated by different layers dominates the computation latency (dashed green line) compared to the low computation requirements of the Lenet model. Accordingly, when the load of requests is low, several Lenet classifications can be processed by a single device without the need for distribution as depicted by the green solid line. Meanwhile, when the online load is high and the memory capacity will not able to handle all requests locally and the system starts to resort to the distribution even when the network is small, e.g. Lenet. 
Similarly, the communication latency increases with the increase of the incoming load, which is explained by the high intermediate data exchange between UAVs and the interferences occurring between participants. The simulation results presented by the blue line \ref{sub-fig:lenet_OULD_latency} is conducted in a network of UAVs characterized by low memory - high computation level. The overall latency of this simulation increases significantly compared to the previous configuration presented by the green line; this is due to the limited memory that restricts the system capacity while increasing the overall latency. The system capacity is defined by the ability of the network to handle a maximum number of parallel classification requests.

The yellow line in the Figure \ref{sub-fig:lenet_OULD_latency} illustrates the optimal latency per request for a network of 15 UAVs deployed in the same target area dimension of 100x100 m where the UAVs are equipped with high memory. We can see that increasing the number of participants contributes to enhance the system capacity. More specifically, the system can handle up to 14 Lenet requests processed autonomously, compared to 10 requests classified locally when involving 10 participants (the green line). Therefore, for the lower loads, only the computation delay impacts the total latency, whereas the communication delay added for distribution dominates the classification time, when the load is high.To summarize, a higher network density contributes to enhance the system capacity to deal with simultaneous classification requests (18 requests when 10 UAVs participates in the collaborative system and 25 when involving 15 devices.). However, when the number of participants is higher, the overall latency increases significantly, which can be explained by the additional overhead resulting from the interference occurring in a dense network as considered in our data rate model (Equation \ref{equ:rho}) based on \cite{8654727}.

\begin{figure}[h]
\centering
\begin{subfigure}[c]{1.0\columnwidth}
		\centering
		\includegraphics[width=\linewidth]{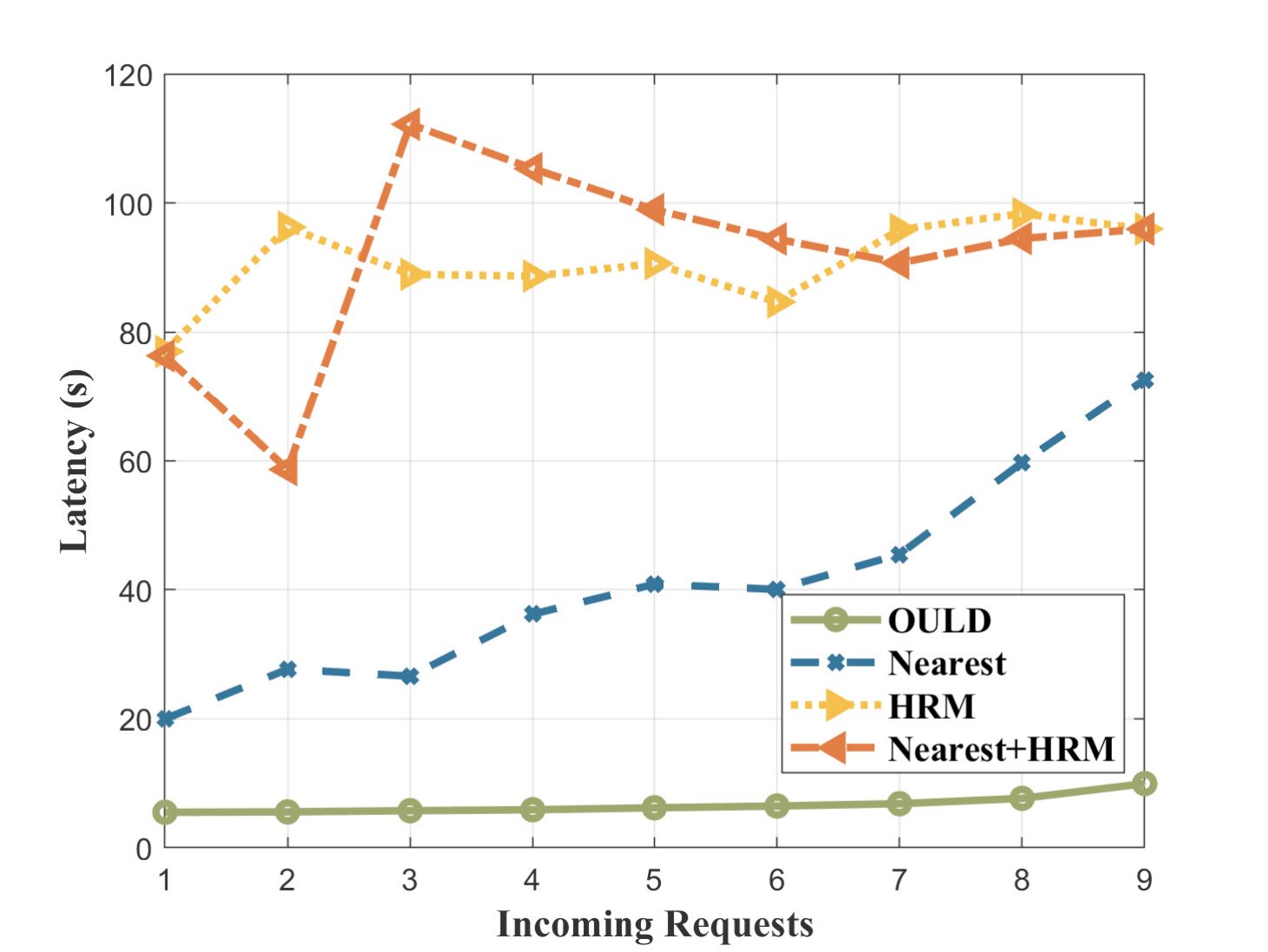}
		\caption{Latency in function of incoming requests}
		\label{fig:comp_OULD_latency}
\end{subfigure}
\hfil{}
\begin{subfigure}[c]{1.0\columnwidth}
		\centering
		\includegraphics[width=\linewidth]{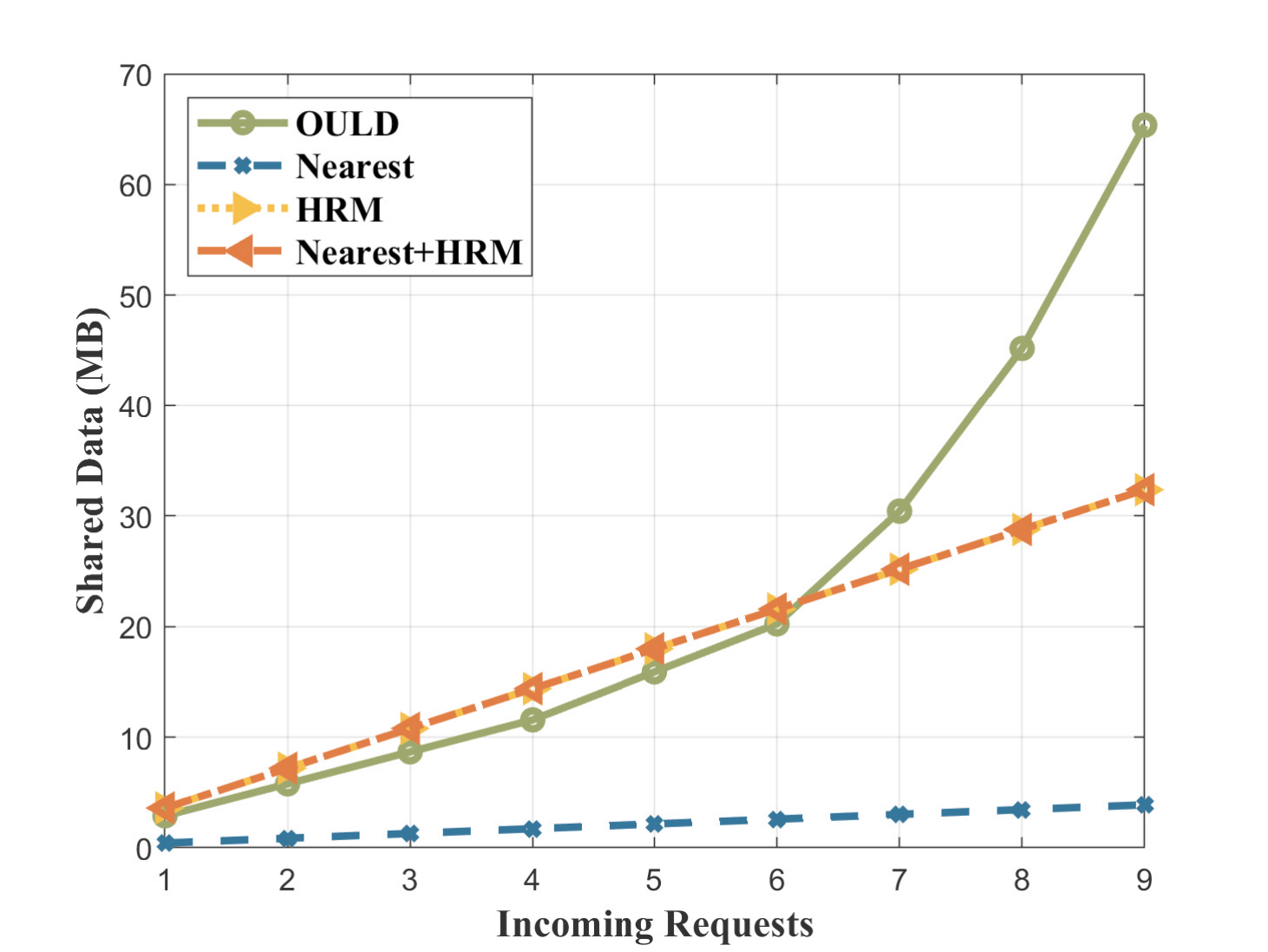}
		\caption{Shared data in function of incoming requests}
		\label{fig:comp_OULD_data}
\end{subfigure}
\caption{Performance comparison of our proposed method OULD VS the heuristic approaches designed for a single network configuration to distribute the inference tasks.}
\label{fig:comp_OULD}
\end{figure}

Figure \ref{sub-fig:lenet_OULD_sharedData} depicts the amount of data exchanged between UAVs to perform distributed inferences following the scenarios simulated in Figure \ref{sub-fig:lenet_OULD_latency}. We can see that a network with low memory UAVs requires high data exchange. This is related to the incapacity to accomplish autonomous classifications and the need to resort to distribution. Meanwhile, a higher network density combined with high memory capabilities contributes to increase significantly the system capacity to compute more layers in a single device and minimize the data sharing. The yellow dotted line depicts the shared data for a network of 15 UAVs with low memory constraints, such as a large amount of data is exchanged in this network to process cooperatively the 12 incoming requests which is the system capacity as shown by the red line in Figure .  This is due to more resource-constrained UAVs are involved in the cooperation process which increases the system capacity by 2 requests compared to the blue dashed line that depicts results for a network of 10 UAVs. After reaching the system capacity the additional incoming request will not be handled in parallel which results in stable shared data as we can see for the green line after 18 requests ( as shown by green line in Figure \ref{sub-fig:lenet_OULD_latency}).

Figures \ref{fig:vgg-10} and \ref{fig:vgg-15} present the performance of our system when receiving requests for classification using VGG-16 network, which is considered as one of the most complex and deep CNN models. Unlike the Lenet model where some requests may be handled locally without distribution, the VGG-16 requires collaborative inference because of its high memory and computation demand. Figure \ref{fig:vgg-10} shows the performance of a network composed of 10 UAVs with different memory  constraints. When the UAVs are endowed with high memory and computation levels (see Figure \ref{sub-fig:vgg-10-HH}), we can notice that the computation highly dominates in the end-to-end latency.

\begin{figure*}[!ht]
     \begin{minipage}[l]{1.0\columnwidth}
\centering
	\begin{subfigure}[b]{\linewidth}
		\centering
		\includegraphics[width=\textwidth]{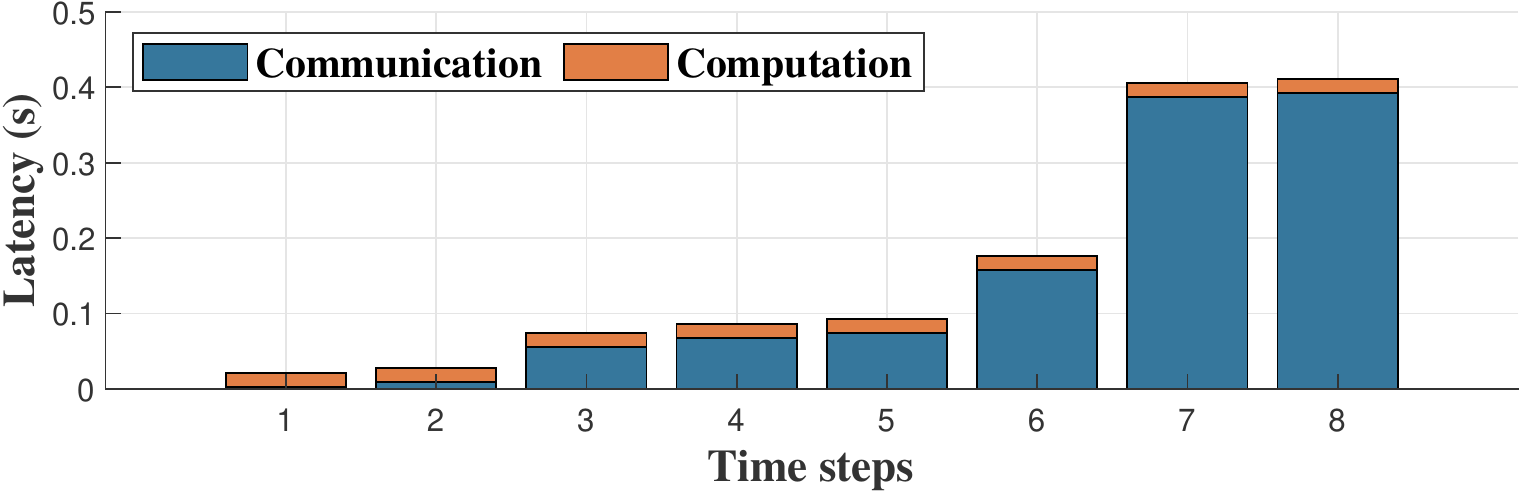}
		\caption{m = 512 MB and c = 9.5 GF}
		\label{sub-fig:lenet-100-HH}
	\end{subfigure}\hfill
	\begin{subfigure}[b]{\linewidth}
		\centering
		\includegraphics[width=\textwidth]{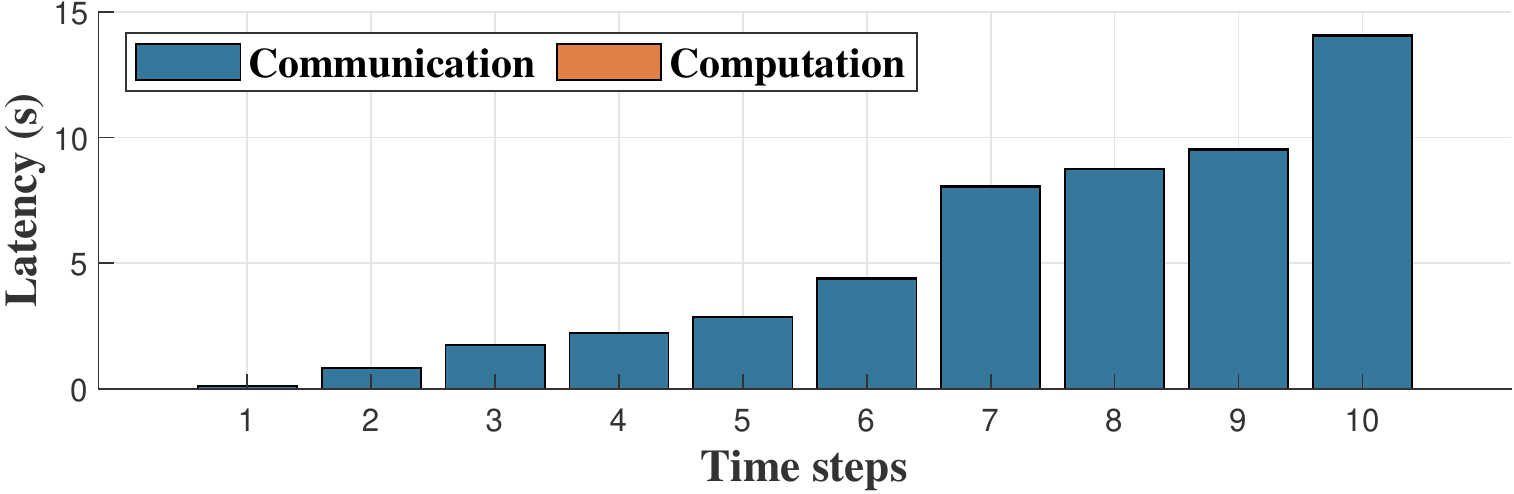}
		\caption{m = 256 MB and c = 9.5 GF}
		\label{sub-fig:lenet-100-LH}
	\end{subfigure}\hfill
	\caption{Optimal decision-making latency of a distributed Lenet's inference over a network of UAVs deployed $100^2 m$ area using mobility prediction.}
	\label{fig:lenet-100}
     \end{minipage}
     \hfill{}
     \begin{minipage}[r]{1.0\columnwidth}
        \centering
	\begin{subfigure}[b]{\linewidth}
		\centering
		\includegraphics[width=\textwidth]{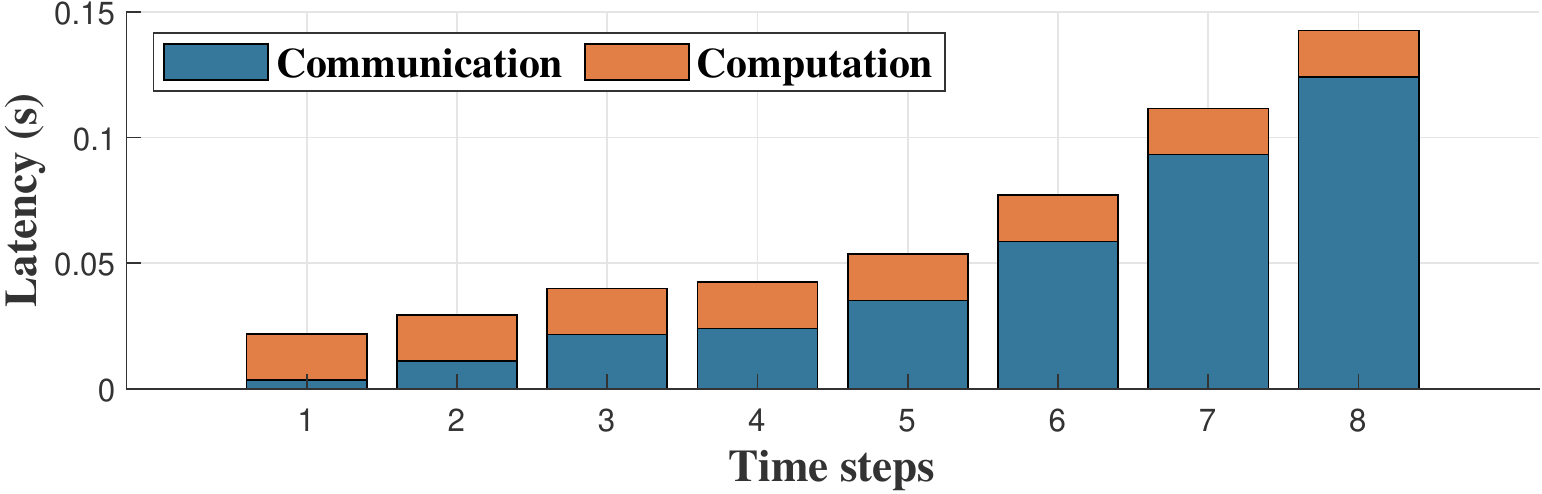}
		\caption{m = 512 MB and c = 9.5 GF}
		\label{sub-fig:lenet-500-HH}
	\end{subfigure}\hfill
	\begin{subfigure}[b]{\linewidth}
		\centering
		\includegraphics[width=\textwidth]{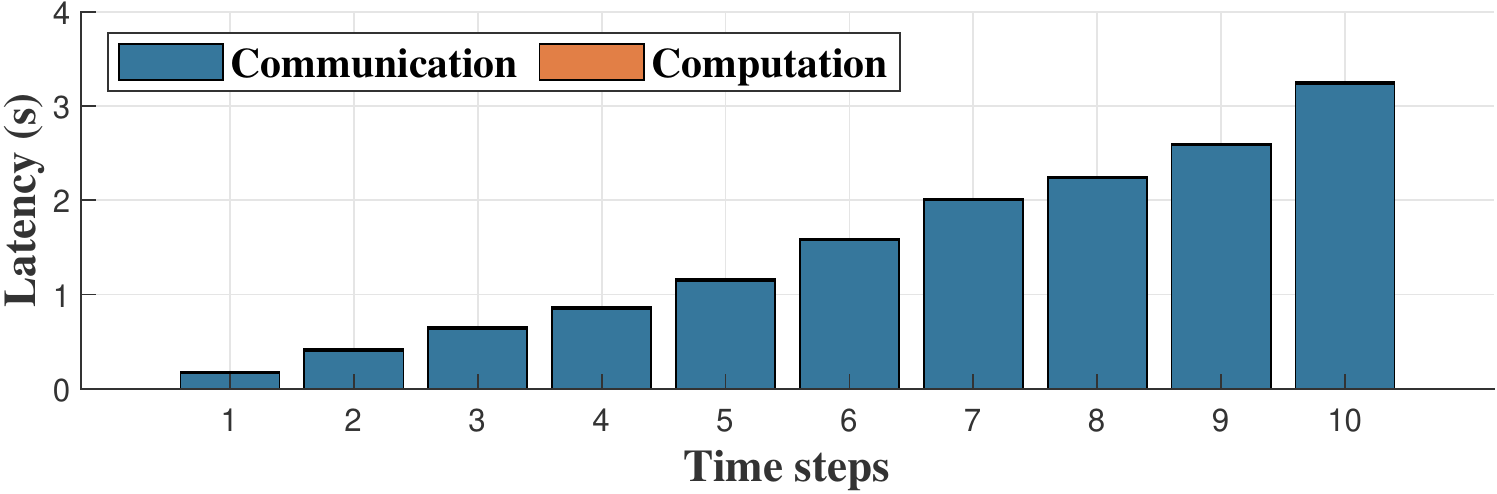}
		\caption{m = 256 MB and c = 9.5 GF}
		\label{sub-fig:lenet-500-LH}
	\end{subfigure}\hfill
	\caption{Optimal decision-making latency of a distributed Lenet's inference over a network of UAVs deployed $500^2 m$ area using mobility prediction.}
	\label{fig:lenet-500}
     \end{minipage}
\end{figure*}

Figure \ref{fig:vgg-15} illustrates the distribution of classification requests over a network composed of 15 UAVs. We can see in Figure \ref{sub-fig:vgg-15-HH} that the optimal latency to process inferences on UAVs with high memory increases when the incoming load is high. When the system reaches its maximum capacity (10 requests), the optimal latency increases tremendously dominated by the communication delay. This huge increase in the communication delay is resulting from involving all the UAVs including the devices with the worse location, which disturbs the overall communication in the network by creating interferences. This scenario is not faced when the number of participants is equal to 10 (see Figure \ref{fig:vgg-10}), due to the small network density and low probability of interference. Otherwise, when the system reaches its maximum capacity, the communication delay becomes greater due to the network density and the interferences created by the low-performance nodes, disturbing the transmission. The performance of our approach when deploying low memory UAV devices is shown in Figure \ref{sub-fig:vgg-15-LH}. We can see that the system capacity to handle online incoming requests is reduced compared to the previous configuration depicted in Figure \ref{sub-fig:vgg-15-HH}, which illustrates again the importance of the memory level endowed with devices in maximizing the number of simultaneous classifications. We also emphasize that the impact of the network density on the end-to-end latency in terms of interference is reduced due to the low number of computed requests. Figure \ref{fig:vgg-data} depicts the amount of data exchanged between UAVs to perform distribution following the scenarios simulated in Figures \ref{fig:vgg-10} and \ref{fig:vgg-15}. In addition to the high computation requirements of VGG-16, we can see that the exchanged data to conduct the cooperative inference is greater than Lenet, which justifies its high inference latency per request. Unlike Lenet , the VGG-16 system deploying UAVs with high memory level exchanges more data than  the one using low memory devices due to its capacity to handle a higher number of requests. The maximum number of VGG requests that can be handled onboard the UAV swarm is smaller compared to Lenet which are respectively 10 and 25 requests, this is due to the high memory and computation requirements of VGG. So as a system deploying a low number of UAVs equipped with low memory capacity achieves low capacity such after only 3 incoming requests start rejecting requests that result in stable shared data as seen for the blue line.

In Figure \ref{fig:comp_OULD}, we show the performance comparison of our OULD method and three heuristic schemes designed to distribute the inference task between UAVs for a single configuration obtained from a fixed time step. In the Nearest heuristic method, the nodes that receive the computing task search among its neighbors the nearest one with enough memory to process the remaining layers. Regarding the High Residual Memory heuristic (HRM), the node selects the UAV in its neighborhood with highest residual memory. The third heuristic combines both the nearest and HRM, where the UAVs requiring cooperation select the nearest UAVs with high residual memory to participate in the cooperative inference. In Figure \ref{fig:comp_OULD_latency}, we plot the incoming request in terms of average latency per request for all methods. We can see that our optimization solution outperforms the heuristic methods. Otherwise, the heuristic method selecting the nearest nodes to transmit its intermediate output performs well compared to the two other heuristic methods. This can be explained by the fact that the nearest nodes possess high data rate links which results in lower average latency per request compared to the method based on memory constraints. Meaning, air-to-air communication has a great impact to determine the optimal distribution policy.

\subsection{OUD-MP evaluation}

In this section, we focus on evaluating the performance of the proposed methodology in section \ref{sec:tailoring}, where we take into consideration the mobility model adopted by the UAVs that describe their movement during the mission and is defined by different locations of UAVs at each point of time. Indeed, since the locations of all UAVs are known during a period of time, we discretize this duration into time steps; each one indicates the new positions of the moving UAVs. In this way, a single distribution policy is adopted during this interval of time;  which enhances the delay performance of the proposed approach by preventing the execution of the optimization problem for each topology variation. Furthermore, solving a one-shot optimization presents an optimal distribution policy, since the system selects UAVs performing different tasks while considering their future locations to ensure the non-disconnection between devices; and consequently guarantee the completeness of the whole classification and the accuracy of the model. This is not the case of the static network configuration, where we execute the optimization for each variation of the system and decisions do not consider possible disconnections or data rate decrease.

\begin{figure*}[!ht]
     \begin{minipage}[l]{1.0\columnwidth}
\centering
	\begin{subfigure}[b]{\linewidth}
		\centering
		\includegraphics[width=\textwidth]{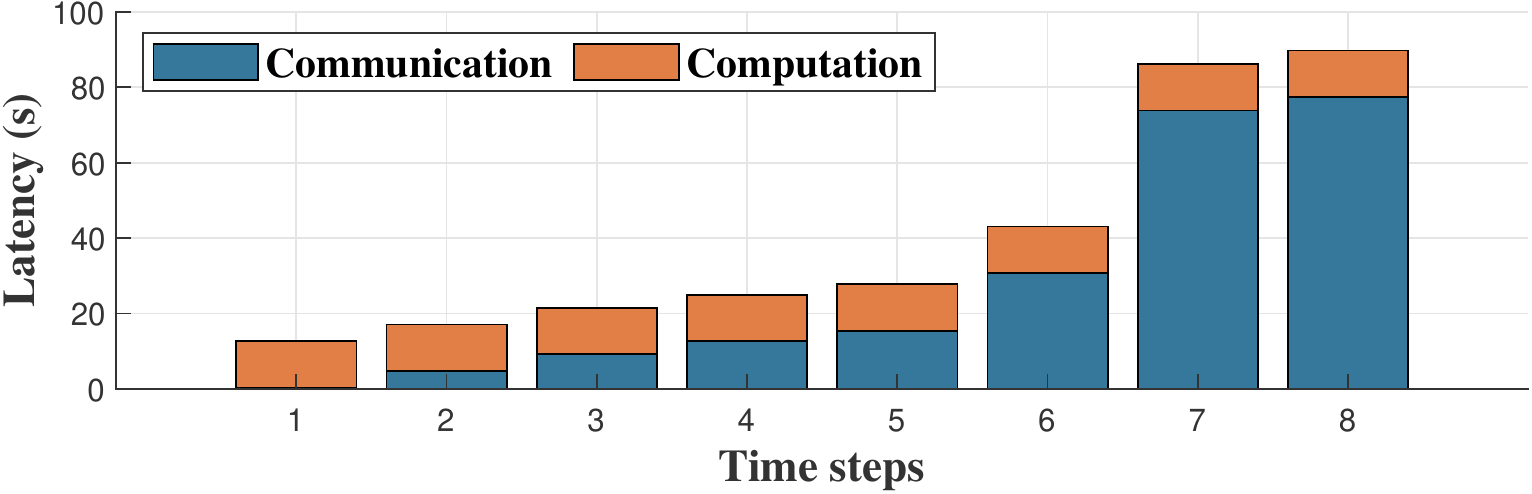}
		\caption{m = 512 MB and c = 9.5 GF}
		\label{sub-fig:vgg-100-HH}
	\end{subfigure}\hfill
	\begin{subfigure}[b]{\linewidth}
		\centering
		\includegraphics[width=\textwidth]{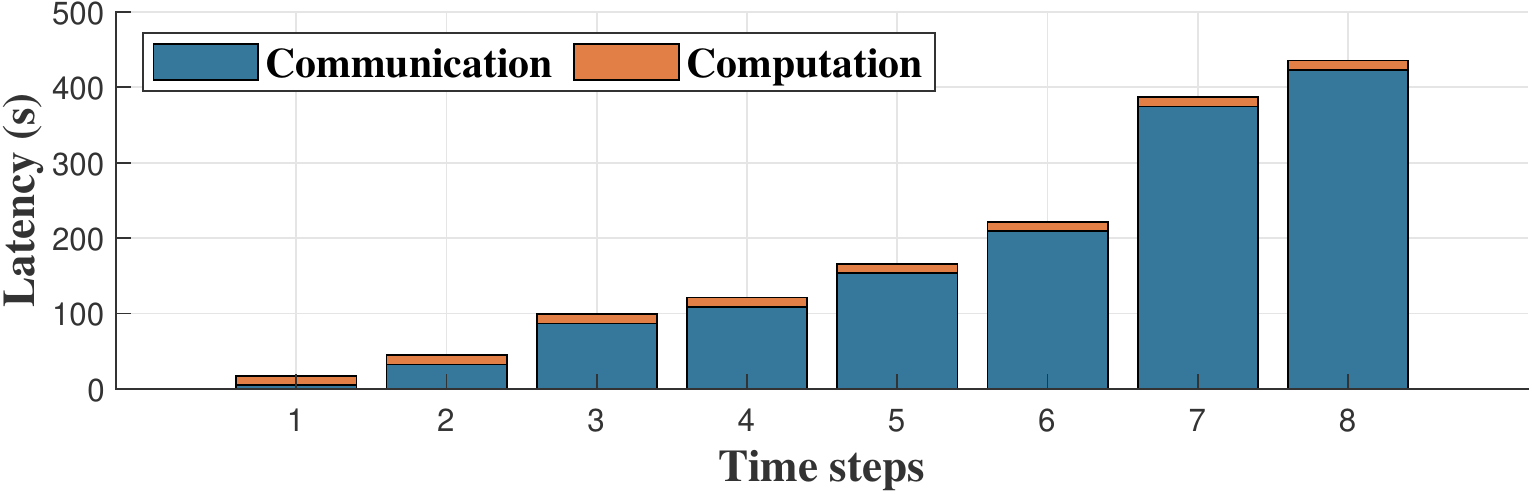}
		\caption{m = 256 MB and c = 9.5 GF}
		\label{sub-fig:vgg-100-LH}
	\end{subfigure}\hfill
	\caption{Optimal decision-making latency of a distributed VGG's inference over a network of UAVs deployed $100^2 m$ area using mobility prediction.}
	\label{fig:vgg-100}
     \end{minipage}
     \hfill{}
     \begin{minipage}[r]{1.0\columnwidth}
       \centering
	\begin{subfigure}[b]{\linewidth}
		\centering
		\includegraphics[width=\textwidth]{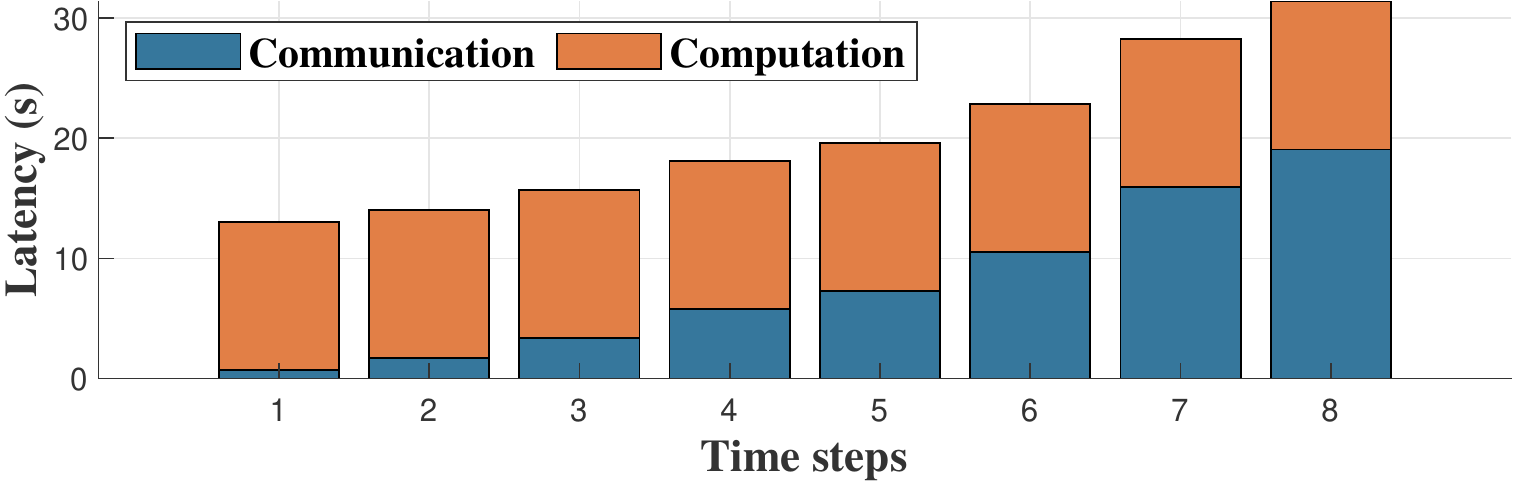}
		\caption{m = 512 MB and c = 9.5 GF}
		\label{sub-fig:vgg-500-HH}
	\end{subfigure}\hfill
	\begin{subfigure}[b]{\linewidth}
		\centering
		\includegraphics[width=\textwidth]{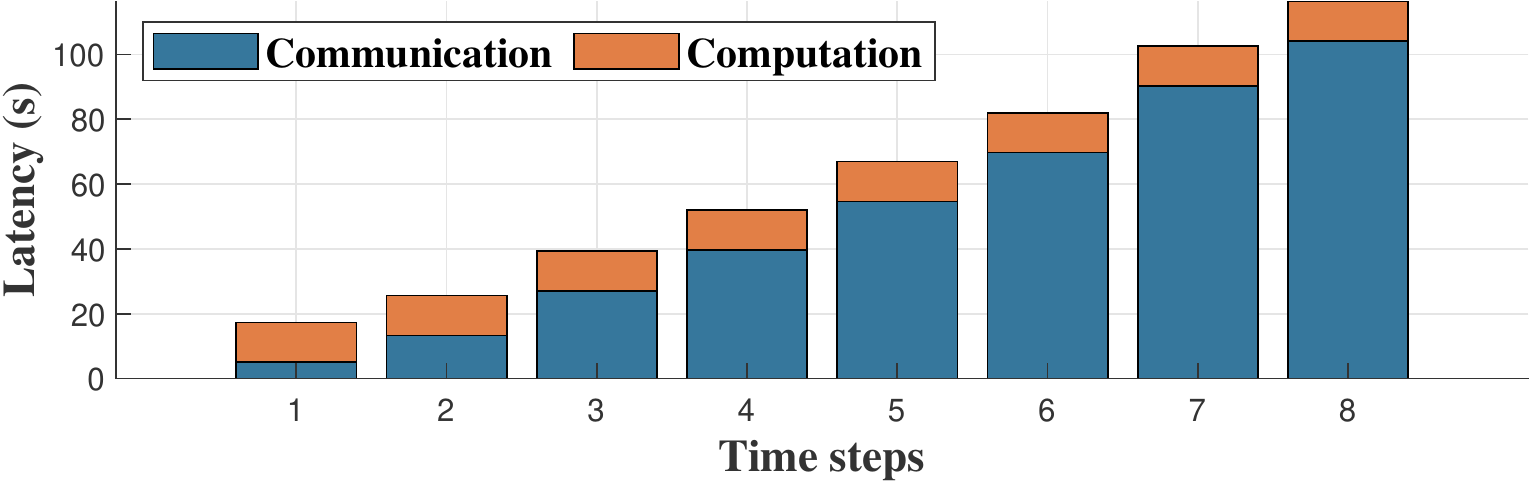}
		\caption{m = 256 MB and c = 9.5 GF}
		\label{sub-fig:vgg-500-LH}
	\end{subfigure}\hfill
	\caption{Optimal decision-making latency of a distributed VGG's inference over a network of UAVs deployed $500^2 m$ area using mobility prediction.}
	\label{fig:vgg-500}
     \end{minipage}
\end{figure*}

Figure \ref{fig:lenet-100} presents the optimal decision-making latency of a UAVs network receiving Lenet requests in a target area of 100 $m^2$, where two memory levels are considered. Our objective through varying the network configuration is to define  a trade-off between the mobility model adopted by the UAVs, their technological characteristics, and the target area dimension. Figure \ref{sub-fig:lenet-100-HH} presents the performance of a UAV network characterized by high memory level. We can see that the latency increases by increasing the number of time steps where the mobility prediction is performed. In fact, for a long mobility period, the obtained policy handles multiple states of the network, which increases the overall latency per request. %This is due to a single policy should handle multiple network topologies. (redundunt no need)
Furthermore, for a network deployed in 100 $m^2$ area, similar or slightly different policies may be used for multiple time steps as shown for steps 3, 4, and 5,  which is due to the low need for cooperation since UAVs with high memory are deployed resulting in a low impact of the topology variation on the inference latency. Figure \ref{sub-fig:lenet-100-LH} shows the performance of a network with a low memory capacity, which means a higher requirement for distribution and consequently a huge data exchange between participants. Particularly, The intermediate output transmission introduces a significant communication delay to the end-to-end latency; which makes the computation delay of such a small model negligible compared to the communication overhead.

\begin{figure*}[!ht]
\begin{subfigure}[c]{1.0\columnwidth}
\centering
\includegraphics[width=\textwidth]{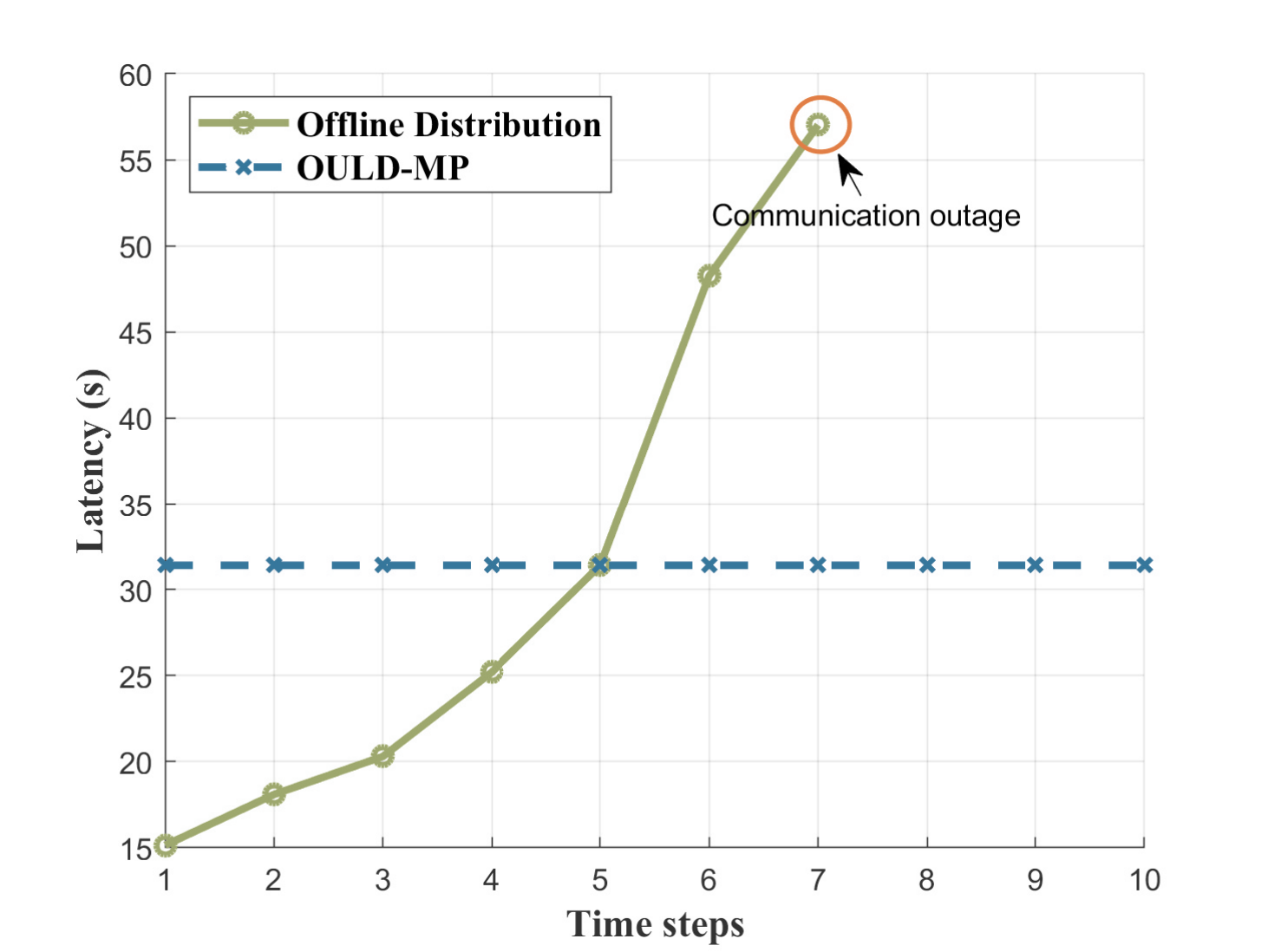}
\caption{Latency in function of incoming requests}
\label{fig:comp_OULD-MP_latency}
\end{subfigure}
\hfill{}
\begin{subfigure}[c]{1.0\columnwidth}
\centering
\includegraphics[width=\textwidth]{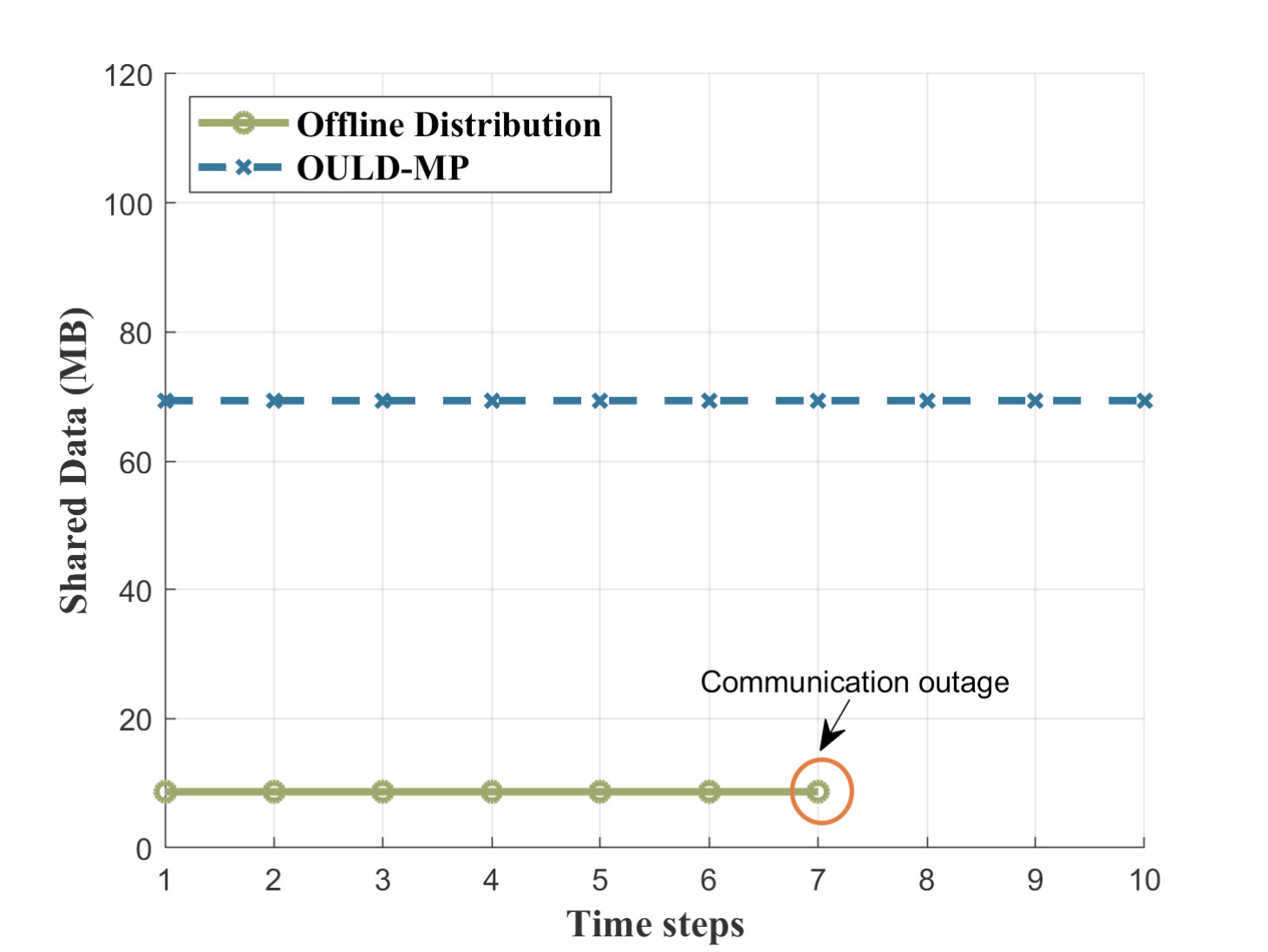}
\caption{Latency in function of incoming requests}
\label{fig:comp_OULD-MP_data}
\end{subfigure}
\caption{Performance comparison of our proposed method OULD-MP using mobility prediction and the offline distribution designed in \cite{disabato2019distributed}.}
\label{fig:comp_OULD-MP}
\end{figure*}

The optimal decision-making latency of a UAV network deployed on a target area of 500 $m^2$ is presented in Figure \ref{fig:lenet-500}, where different UAVs properties are examined. Figure \ref{sub-fig:lenet-500-HH} shows the performance of a network of UAVs equipped with high memory capability. We can see that the end-to-end latency increases through time steps of the mobility prediction, which is related to the fact that a single policy should handle multiple topologies observed during this period of time. 

Unlike the previous network configuration evaluated in Figure \ref{sub-fig:lenet-100-HH}, a single distribution policy cannot be found for multiple time steps due to the wide target area resulting in a high topology variation, which increases the impact on the inference time. In addition, we can notice a lower overall latency compared to the same network deployed in a 100 $m^2$ area and evaluated in Figure \ref{sub-fig:lenet-100-HH}. This can be explained by the low interference between UAVs, as they are deployed in a wide area of 500 $m^2$. Figure \ref{sub-fig:lenet-500-LH} shows the performance of the network of UAVs equipped with a low memory capacity. We can see that the communication delay dominates compared to the computation latency, due to the memory limitation, the need for distribution, and accordingly the high data exchange.

Figures \ref{fig:vgg-100} and \ref{fig:vgg-500} depict the optimal decision-making latency when distributing VGG-16 inferences over UAVs deployed respectively in 100 $m^2$ and 500 $m^2$ areas. The performance of the network composed of UAVs characterized by a high memory capability is shown in Figure \ref{sub-fig:vgg-100-HH}. This Figure shows that the overall latency increases over time steps as different policies are required for each mobility prediction duration. In this scenario, the overall latency increases over time to handle multiple topology variations. We note that adopting single policy as done for Lenet cannot be applied for VGG due to its large depth and the low memory capacity of the deployed UAVs. Figure \ref{fig:vgg-500} shows the performance of a network of UAVs deployed in a target area of 500 $m^2$. In Figure \ref{sub-fig:vgg-500-HH}, we can clearly notice that the computation time hugely dominates the overall latency when the UAVs are equipped with high memory capability. Meaning, an efficient inference distribution is designed, particularly for low mobility prediction duration, since the overall latency is spent on computation in addition to high distribution demand due to resource-constrained UAVs. 

In Figure \ref{fig:comp_OULD-MP}, we show the performance comparison of our proposed method OULD-MP that uses the mobility prediction and the reference method in \cite{disabato2019distributed}, where the offline distribution of multiple models is suggested. In Figure \ref{fig:comp_OULD-MP_latency}, our OULD-MP shows stable average latency per request during the 10-time steps where the UAVs are moving in the targeted region. This is due to the mobility prediction used in our proposed method to ensure optimal distribution for a bounded duration of time. The method of \cite{disabato2019distributed} shows good results for the first 5 time slots compared to OULD-MP since the optimal policy deployed does not take into consideration the new location of UAVs, such as the involved nodes in the distribution are moving away from each other, which results in an outage of communication in time step 7. Consequently, the incoming requests can not be processed collaboratively and may be rejected.

\begin{figure}[t]
\begin{center}
\includegraphics[scale=0.6]{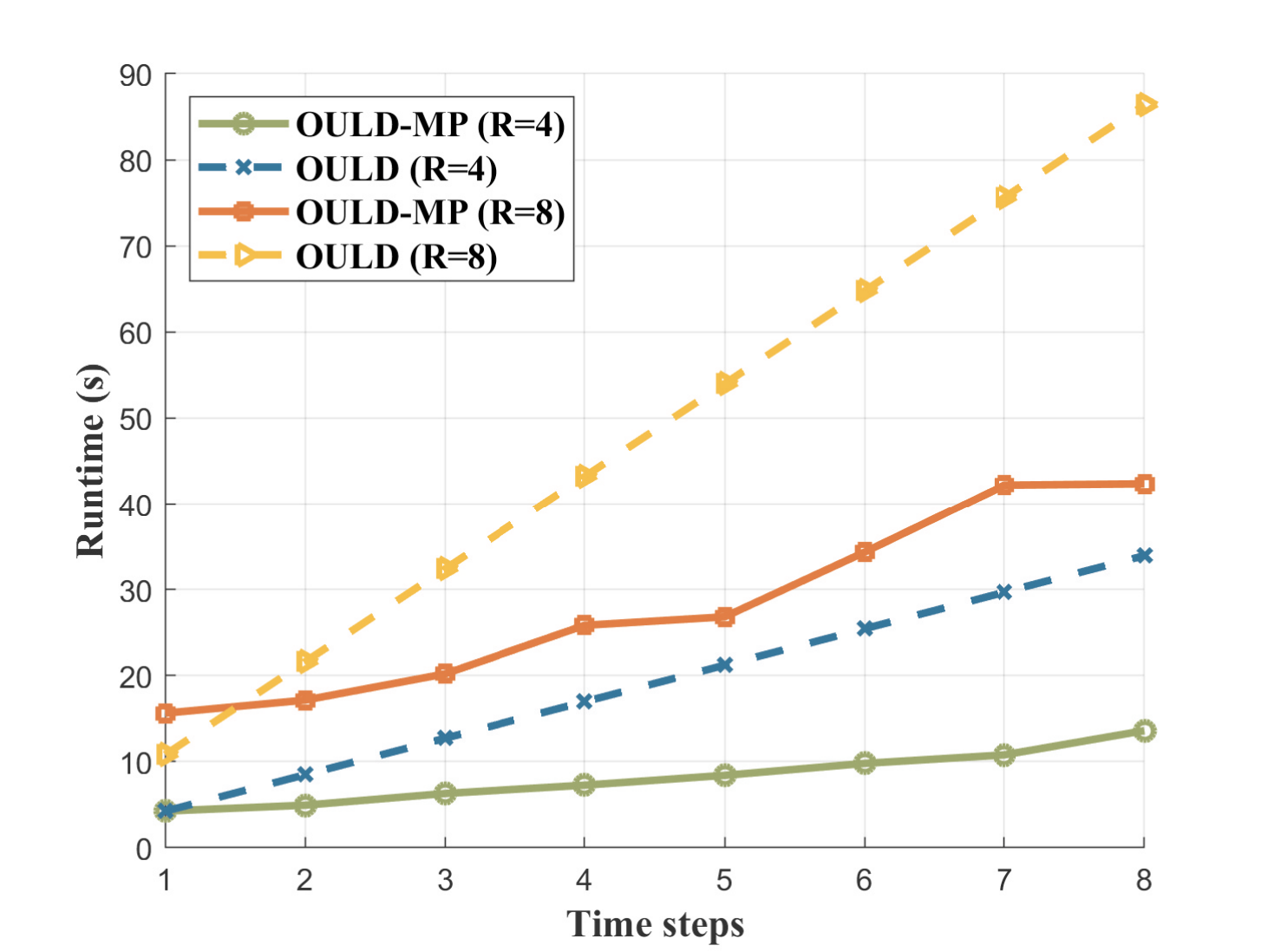}
\caption{Runtime comparison of both OULD and ML-OULD with mobility prediction versus the time steps for 4 and 8 incoming requests at each time step.}
\label{fig:lenet_runtime}
\end{center}
\end{figure}

\subsection{Complexity analysis}

The computational complexity of an optimization problem depends, in general, on the problem formulation and the technique that is used for solving it. In some algorithms, the complexity can be measured by the time that the CPU needs to run the algorithm, others approximate the computational complexity by the number of constraints and the nested loops in the objective function per run, which can be written as $O(.)$. In both approaches proposed, we have the same constraints. The difference resides in the objective functions, whose complexity is equal to  $O(RN^2(M-1))$ for the static configuration OULD and $O(TRN^2(M-1))$ for OULD-MP. However, when dealing with the static configuration, the optimization needs to be re-run for each network change (e.g. topology variation, new incoming requests, disconnection of UAVs, etc.). Hence, the runtime of a single problem needs to be multiplied by $T$ that denotes the number of time slots where the system changes. Meanwhile, OULD-MP is a one shot optimization. On these basis, comparing both problems' runtime complexity is not straightforward. Therefore, we measured the time that the CPU requires to run the two approaches, following the same parameters. Figure \ref{fig:lenet_runtime} presents a comparison between OULD and OULD-MP in terms of runtime, for different loads of concurrent requests and when increasing the number of time steps. %We note that we tested the time complexity of the optimizations on a computer, having the following characteristics: core i7 and 16 GB RAM while the main results of VGG16 simulation are performed on an HPC as described it the beginning of section IV. 
We can see that for different configurations, OULD-MP covering the mobility of devices presents a lower runtime than OULD approach, where we need to run the simulation each time the network changes. Hence, the mobility study adds not only an overview about the status of the system in the next time slots to avoid any disconnection scenario and establish transmissions between devices with expected higher data rates, but also presents less complexity and better runtime.

\subsection{Results Summary}

We used in our simulation Lenet and VGG-16 due to their different complexities and computation requirements in order to study the tradeoff between the number of requests that could be processed in parallel, the number of UAVs available to perform collaborative inference as well as their computation capabilities. In addition, we considered the impact of the wireless communication between the UAVs on the end-to-end latency through a data rate model from [38] including the interferences that occur in the network. %\hl{Although multiple parameters such as air traffic control, climb rate, reliability were not covered in this study  since we were supposed to work with a set of UAVs deployed in the targeted area for surveillance purposes. These UAVs coordinate their movement following the reference point mobility model in order to cover all-region.} 
Although multiple parameters such as air traffic control, climb rate, reliability were not covered in this study, we decided to cover two critical parameters (average latency and shared data) due to the characteristics of our UAVs system. This later is composed of a set of UAVs deployed in the targeted area for surveillance purposes such as all the UAVs coordinate their movement following the reference point mobility model in order to cover the target surface.

As shown in our study, some UAVs with limited computation and memory capabilities are not able to process an entire request of high complexity as VGG-16 or multiple incoming simple requests as Lenet. Our OULD method showed efficient results to process collaboratively multiple incoming requests whatever their type is instead of using a server to remotely process the collected data. The use of a server introduces multiple drawbacks, especially for online systems that require real-time processing, such as the high latency spent on communicating the data to the server and the complex deployment of the system. Furthermore, When the mobility model followed by the UAVs participating in the collaborative inference process is non-homogeneous, the link quality between UAVs will change or can eventually be lost. For this reason, our OULD method should be executed each time step for each network configuration, which makes it impractical, non optimal on overall throughout all time steps and introducing a computational complexity to the optimization. To solve this issue, we proposed the use of mobility prediction (ML-OULD) where the positions of all UAVs are known for some time steps that allow performing distribution, while taking into consideration the future positions of UAVs and avoiding periodic execution of OULD method. As a result, our ML-OULD method showed efficient performance in terms of decision making latency while considering time variation of the wireless channel between the UAVs.

%=====================================================================
\section{Conclusion}\label{conclusion}
A distributed CNN inference method is proposed in this paper in order to fit the CNN memory requirements to the resource-constrained UAVs and allow performing on-site classification instead of delegating the decision-making to a server.%The aim is to avoid delay and bandwidth consumed for data transmission into the server which is inefficient especially for real-time systems due to narrow band and highly variant A2G channel quality.
The distribution scheme in this paper is formulated as an optimization problem, where the objective function consists of minimizing the decision-making latency. In addition, the mobility model followed by the UAVs and their varying topology are well studied in our optimization due to its impact on the decision-making latency. 

Furthermore, a mobility prediction is considered in the tailored model to prevent multiple executions of the optimization problem for each topology variation. Our simulation unveiled different parameters that should be present to achieve an efficient distribution of CNNs, including the density
of UAVs, their capacities, and the mobility model. However, the optimization is characterized by its complexity, which makes finding the optimal solution extremely challenging. For this reason, we plan in future works to use reinforcement learning as an alternative to the optimization as it showed a high capacity to pursue the optimal solution in resource allocation scenarios.

% if have a single appendix:
%\appendix[Proof of the Zonklar Equations]
% or
%\appendix  % for no appendix heading
% do not use \section anymore after \appendix, only \section*
% is possibly needed

% use appendices with more than one appendix
% then use \section to start each appendix
% you must declare a \section before using any
% \subsection or using \label (\appendices by itself
% starts a section numbered zero.)
%

% Can use something like this to put references on a page
% by themselves when using endfloat and the captionsoff option.
\ifCLASSOPTIONcaptionsoff
  \newpage
\fi

% trigger a \newpage just before the given reference
% number - used to balance the columns on the last page
% adjust value as needed - may need to be readjusted if
% the document is modified later
%\IEEEtriggeratref{8}
% The "triggered" command can be changed if desired:
%\IEEEtriggercmd{\enlargethispage{-5in}}

% references section

% can use a bibliography generated by BibTeX as a .bbl file
% BibTeX documentation can be easily obtained at:
% http://mirror.ctan.org/biblio/bibtex/contrib/doc/
% The IEEEtran BibTeX style support page is at:
% http://www.michaelshell.org/tex/ieeetran/bibtex/
%\bibliographystyle{IEEEtran}
% argument is your BibTeX string definitions and bibliography database(s)
%\bibliography{IEEEabrv,../bib/paper}
%
% <OR> manually copy in the resultant .bbl file
% set second argument of \begin to the number of references
% (used to reserve space for the reference number labels box)
\bibliographystyle{IEEEtran}
\bibliography{references}

% biography section
% 
% If you have an EPS/PDF photo (graphicx package needed) extra braces are
% needed around the contents of the optional argument to biography to prevent
% the LaTeX parser from getting confused when it sees the complicated
% \includegraphics command within an optional argument. (You could create
% your own custom macro containing the \includegraphics command to make things
% simpler here.)
%\begin{IEEEbiography}[{\includegraphics[width=1in,height=1.25in,clip,keepaspectratio]{mshell}}]{Michael Shell}
% or if you just want to reserve a space for a photo:
\vspace{-0.8cm}
\begin{IEEEbiography}[{\includegraphics[width=1in,height=1.25in,clip,keepaspectratio]{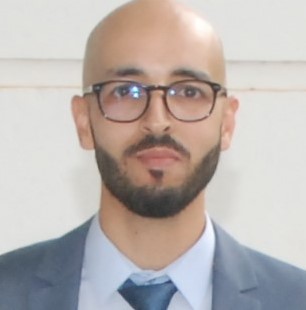}}]{Mohammed Jouhari}
received the B.Sc. degree in physics and the M.Sc. degree in signals processing and telecommunication from the Faculty of Sciences, Mohammed V University, Rabat, Morocco, in 2011 and 2013, respectively. Dr. Jouhari obtained his Ph.D. from Ibn Tofail University, Kenitra, Morocco in 2019. He is currently a Postdoc Fellow at the Computer Science and Engineering Department in Qatar University, Qatar. His research interest include wireless communication, underwater acoustic sensor networks, and distributed machine learning. He is also interested in the internet of things and deep reinforcement learning.
\end{IEEEbiography}
\vspace{-0.5cm}
\begin{IEEEbiography}[{\includegraphics[width=1in,height=1.25in,clip,keepaspectratio]{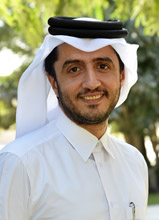}}]{Abdulla Khalid Al-Ali}
, Ph.D. obtained his Master degree in Software Design Engineering and PhD degree in Computer Engineering from Northeastern University in Boston, MA, USA in 2008 and 2014, respectively. He is an active researcher in Cognitive Radios for smart cities and vehicular ad-hoc networks (VANETs). He has published a number of peer-reviewed papers in journals and conferences. He has been awarded the Platinum medal in the Educational Excellence Day Prize for Ph.D. holders in 2015. Dr. Abdulla is currently head of the Computer Science and Engineering Department at the College of Engineering in Qatar University.
\end{IEEEbiography}
\vspace{-1cm}
\begin{IEEEbiography}[{\includegraphics[width=1in,height=1.25in,clip,keepaspectratio]{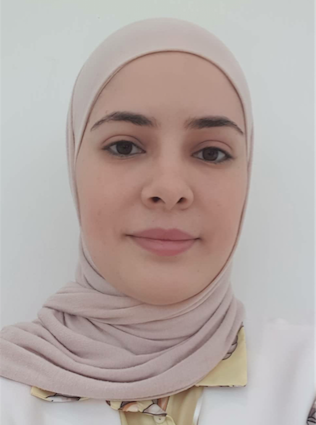}}]{Emna Baccour}
received the Ph.D. degree in computer Science from the University of Burgundy, France, in 2017. She was a postdoctoral fellow at Qatar University on a project covering the interconnection networks for massive data centers and then on a project covering video caching and processing in mobile edge computing networks. She currently holds a postdoctoral position at Hamad Ben Khalifa University. Her research interests include data center networks, cloud computing, green computing and software defined networks as well as distributed systems. She is also interested in edge networks and mobile edge caching and computing.
\end{IEEEbiography}
\vspace{-1cm}
\begin{IEEEbiography}[{\includegraphics[width=1in,height=1.25in,clip,keepaspectratio]{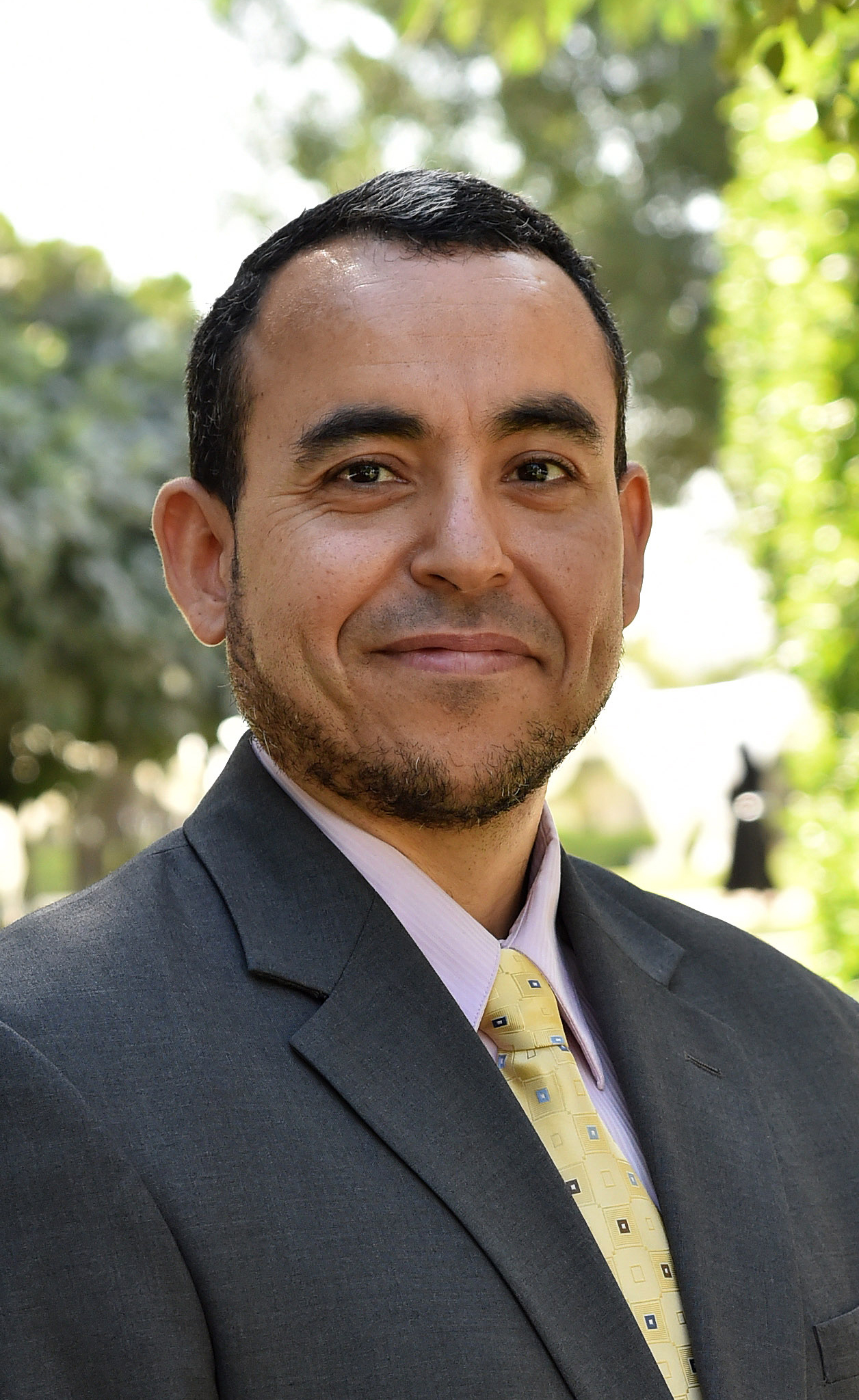}}]{Amr Mohamed}
(S' 00, M' 06, SM' 14) received his M.S. and Ph.D. in electrical and computer engineering from the University of British Columbia, Vancouver, Canada, in 2001, and 2006 respectively. He has worked as an advisory IT specialist in IBM Innovation Centre in Vancouver from 1998 to 2007, taking a leadership role in systems development for vertical industries. He is currently a professor in the college of engineering at Qatar University and the director of the Cisco Regional Academy. He has over 25 years of experience in wireless networking research and industrial systems development. He holds 3 awards from IBM Canada for his achievements and leadership, and 4 best paper awards from IEEE conferences. His research interests include wireless networking, and edge computing for IoT applications. Dr. Amr Mohamed has authored or co-authored over 160 refereed journal and conference papers, textbook, and book chapters in reputable international journals, and conferences. He is serving as a technical editor for the journal of internet technology and the international journal of sensor networks. He has served as a technical program committee (TPC) co-chair for workshops in IEEE WCNC'16. He has served as a co-chair for technical symposia of international conferences, including Globecom'16, Crowncom'15, AICCSA'14, IEEE WLN'11, and IEEE ICT'10. He has served on the organization committee of many other international conferences as a TPC member, including the IEEE ICC, GLOBECOM, WCNC, LCN and PIMRC, and a technical reviewer for many international IEEE, ACM, Elsevier, Springer, and Wiley journals.
\end{IEEEbiography}
\vspace{-1cm}
\begin{IEEEbiography}[{\includegraphics[width=1in,height=1.25in,clip,keepaspectratio]{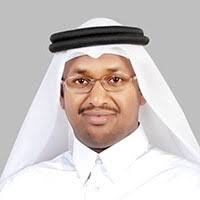}}]{Aiman Erbad}
is an Associate Professor at the College of Science and Engineering at Hamad Bin Khalifa University (HBKU). Dr. Erbad obtained a Ph.D. in Computer Science from the University of British Columbia (Canada), and a Master of Computer Science in Embedded Systems and Robotics from the University of Essex (UK). Dr. Erbad received the Platinum award from H.H. The Emir Sheikh Tamim bin Hamad Al Thani at the Education Excellence Day 2013 (Ph.D. category). Dr. Erbad received the 2020 best research paper award from the Computer Communications journal, IWCMC 2019 best paper award, and IEEE CCWC 2017 best paper award. Dr. Erbad is an editor in KSII Transactions on Internet and Information Systems and was a guest editor in IEEE Networks. Dr. Erbad research interests span cloud computing, edge computing, IoT, private and secure networks, and multimedia systems.
\end{IEEEbiography}
\vspace{-1cm}
\begin{IEEEbiography}[{\includegraphics[width=1in,height=1.25in,clip,keepaspectratio]{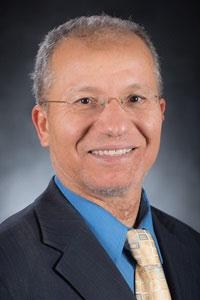}}]{Mohsen Guizani}
(S'85–M'89–SM'99–F'09) received the B.S. (with distinction) and M.S. degrees in electrical engineering, the M.S. and Ph.D. degrees in computer engineering from Syracuse University, Syracuse, NY, USA, in 1984, 1986, 1987, and 1990, respectively. He is currently a Professor at the Computer Science and Engineering Department in Qatar University, Qatar. Previously, he served in different academic and administrative positions at the University of Idaho, Western Michigan University, University of West Florida, University of Missouri-Kansas City, University of Colorado-Boulder, and Syracuse University. His research interests include wireless communications and mobile computing, computer networks, mobile cloud computing, security, and smart grid. He is currently the Editor-in-Chief of the IEEE Network Magazine, serves on the editorial boards of several international technical journals and the Founder and Editor-in-Chief of Wireless Communications and Mobile Computing journal (Wiley). He is the author of nine books and more than 500 publications in refereed journals and conferences. He guest edited a number of special issues in IEEE journals and magazines. He also served as a member, Chair, and General Chair of a number of international conferences. Throughout his career, he received three teaching awards and four research awards. He also received the 2017 IEEE Communications Society WTC Recognition Award as well as the 2018 AdHoc Technical Committee Recognition Award for his contribution to outstanding research in wireless communications and Ad-Hoc Sensor networks. He was the Chair of the IEEE Communications Society Wireless Technical Committee and the Chair of the TAOS Technical Committee. He served as the IEEE Computer Society Distinguished Speaker and is currently the IEEE ComSoc Distinguished Lecturer. He is a Fellow of IEEE and a Senior Member of ACM.
\end{IEEEbiography}
\vspace{-1cm}
\begin{IEEEbiography}[{\includegraphics[width=1in,height=1.25in,clip,keepaspectratio]{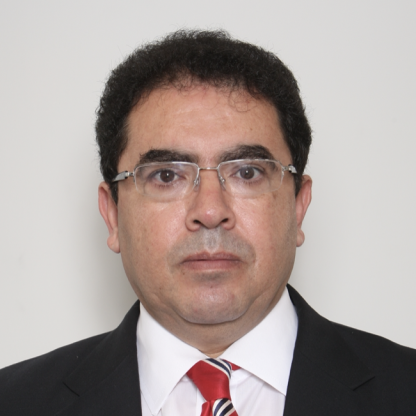}}]{Mounir Hamdi}
 received the B.S. degree (Hons.) in electrical engineering (computer engineering) from the University of Louisiana, in 1985, and the M.S. and Ph.D. degrees in electrical engineering from the University of Pittsburgh, in 1987 and 1991, respectively. He was a Chair Professor and a Founding Member of The Hong Kong University of Science and Technology (HKUST), where he was the Head of the Department of Computer Science and Engineering. From 1999 to 2000, he held visiting professor positions at Stanford University and the Swiss Federal Institute of Technology. He is currently the Founding Dean of the College of Science and Engineering, Hamad Bin Khalifa University (HBKU). His area of research is in high-speed wired/wireless networking, in which he has published more than 360 publications, graduated more 50 M.S./Ph.D. students, and awarded numerous research grants. In addition, he has frequently consulted for companies and governmental organizations in the USA, Europe, and Asia. He is a Fellow of the IEEE for his contributions to design and analysis of high-speed packet switching, which is the highest research distinction bestowed by IEEE. He is also a frequent keynote speaker in international conferences and forums. He is/was on the editorial board of more than ten prestigious journals and magazines. He has chaired more than 20 international conferences and workshops. In addition to his commitment to research and academic/professional service, he is also a dedicated teacher and a quality assurance educator. He received the Best 10 Lecturer Award and the Distinguished Engineering Teaching Appreciation Award from HKUST. He is frequently involved in higher education quality assurance activities as well as engineering programs accreditation all over the world.
\end{IEEEbiography}
\vskip 0pt plus -1fil

% You can push biographies down or up by placing
% a \vfill before or after them. The appropriate
% use of \vfill depends on what kind of text is
% on the last page and whether or not the columns
% are being equalized.

%\vfill

% Can be used to pull up biographies so that the bottom of the last one
% is flush with the other column.
%\enlargethispage{-5in}

% that's all folks
\end{document}